\documentclass[a4paper,11pt]{article}
\pdfoutput=1 

\usepackage{jheppub} 

\reversemarginpar

\usepackage{graphicx}
\usepackage{amsmath}
\usepackage{hyperref}
\usepackage{pdfpages}
\usepackage[section]{placeins}
\usepackage{epstopdf}
\usepackage{tensor}
\usepackage[colorinlistoftodos]{todonotes}
\usepackage{epsfig}
\usepackage{graphicx,color}
\usepackage{cancel}
\usepackage[utf8]{inputenc}
\usepackage{simpler-wick}
\usepackage{comment}
\usepackage{slashed}
\usepackage{tcolorbox}
\usepackage{xcolor}
\usepackage{amssymb}
\usepackage{multirow}
\usepackage{hhline}
\usepackage{tabularx}
\usepackage[export]{adjustbox}
\usepackage{pgfplots}
\usepackage{xcolor}
\pgfplotsset{compat=1.18}
\usepackage{etoolbox}  
\usepackage[titletoc]{appendix}
\appto\appendix{\numberwithin{equation}{section}}
\usepackage{bm}
\usepackage{subcaption}
\usepackage{float}
\usepackage{lineno}   

\newcommand{\La}{\Lambda}
\newcommand{\maplambda}{\texttt{MAPFF1.0\_Lambda}}

\title{First next-to-next-to-leading-order extraction of fragmentation functions for \boldmath{\(\Lambda\)} hyperons}

\author[a]{Valerio Bertone,}
\author[b]{Alessia Bongallino,}
\author[c]{Amedeo Chiefa,}
\author[b]{Miguel G. Echevarría,}
\author[b,d]{Gunar Schnell}
\affiliation[a]{Université Paris-Saclay, CEA, IRFU, 91191 Gif-sur-Yvette, France}
\affiliation[b]{Department of Physics \& EHU Quantum Center, University of the Basque Country (EHU), P.O.~Box 644, 48080 Bilbao, Spain}
\affiliation[c]{The Higgs Centre for Theoretical Physics, University of Edinburgh, JCMB, KB, Mayfield Rd, Edinburgh EH9 3JZ, Scotland}
\affiliation[d]{IKERBASQUE, Basque Foundation for Science, Plaza Euskadi 5, 48009 Bilbao, Spain}
\emailAdd{valerio.bertone@cea.fr}
\emailAdd{alessia.bongallino@ehu.eus}
\emailAdd{amedeo.chiefa@ed.ac.uk}
\emailAdd{miguel.garciae@ehu.eus}
\emailAdd{gunar.schnell@desy.de}

\abstract{
We present \maplambda, the first global analysis at next-to-next-to-leading order in perturbative QCD of the collinear unpolarised fragmentation functions of $\Lambda$ hyperons. The fit is based on data from single-inclusive electron-positron annihilation, and from both neutral-current and --- for the first time --- charged-current semi-inclusive deep-inelastic scattering. We have adopted a statistical framework based on Monte Carlo sampling and parametrised fragmentation functions in terms of a neural network.
The fragmentation function set comprises a total of seven independent parton flavours, allowing for the first independent determination of valence-quark distributions. 
Our analysis offers new insights into the hadronisation mechanism of strange baryons and establishes a baseline for future phenomenological and experimental investigations.
}

\begin{document}
\maketitle
\section{Introduction}

In high-energy collisions, fragmentation functions (FFs)~\cite{Collins:1981uw, Collins:2011zzd, Metz:2016swz} describe the non-perturbative hadronisation of a quark or a gluon into an observed hadron. In particular, collinear FFs carry information about the fraction $z$ of longitudinal momentum that the hadron has inherited from the parent parton. The universality of FFs allows for their extraction from different processes~\cite{Collins:1989gx}. Processes sensitive to FFs include single-inclusive annihilation (SIA), semi-inclusive deep-inelastic scattering (SIDIS), and inclusive hadron production in proton-proton ($pp$) collisions. Collinear FFs play an essential role in that they act as a parton-flavour tagger in processes involving final-state hadrons. Furthermore, in transverse-momentum-dependent (TMD) analyses, TMD FFs are perturbatively matched onto collinear FFs. Hence, a precise knowledge of collinear FFs is crucial for reliable extractions of TMD distributions.

Among the various FFs, the most extensively studied are the collinear unpolarised ones.
Their determination has improved significantly in recent years, driven by the availability of new experimental measurements and by increasingly accurate calculation of partonic cross sections. In particular, while next-to-next-to-leading-order (NNLO) corrections to SIA have been known for a long time~\cite{Rijken:1996vr, Rijken:1996ns, Rijken:1996npa, Blumlein:2006rr, Mitov:2006wy}, NNLO corrections to SIDIS~\cite{Bonino:2024qbh, Goyal:2023zdi, Bonino:2025qta, Goyal:2024emo, Dong:2026eas, Goyal:2026ccx} and $pp$ collisions~\cite{Czakon:2025yti} have become available only recently.
Beyond NNLO, recent works include specific calculations for SIA~\cite{He:2025hin} and SIDIS~\cite{Dong:2026uvw}, the latter based on a transverse-momentum subtraction scheme \cite{Gao:2026tnd}, towards future next-to-next-to-next-to-leading-order descriptions.

Most extractions~\cite{Kniehl:2000fe, Kretzer:2000yf, Bourhis:2000gs, Kretzer:2001pz, Albino:2005me, Albino:2005gd,deFlorian:2007aj, Hirai:2007cx, Anderle:2015lqa, deFlorian:2014xna,  Sato:2016wqj, Bertone:2017tyb,  deFlorian:2017lwf,  Soleymaninia:2018uiv,  Abdolmaleki:2021yjf, Khalek:2021gxf,  Moffat:2021dji, AbdulKhalek:2022laj, Borsa:2022vvp, Gao:2024dbv, Li:2024etc} focus on the lightest and most abundantly produced hadrons, namely pions and kaons. FFs for other species have been far less studied due to the limited availability and precision of experimental data.
Moreover, hadronisation into mesons qualitatively differs from that into baryons.
Therefore, improving the determination of FFs for less known hadrons is essential for achieving a more complete understanding of the hadronisation process.
The $\La$ hyperon, the lightest of strange baryons, will be the focus of this work.

The first determination of collinear FFs for $\La$ production was obtained by the DSV group~\cite{deFlorian:1997zj}. The fit was performed for both unpolarised and helicity-dependent FFs, at next-to-leading order (NLO), and was based on SIA data only.
This extraction assumed that all light quark and antiquark FFs were equal. Later, a more refined NLO analysis of the unpolarised FF was performed by the AKK group~\cite{Albino:2005mv} (dubbed AKK08), which extended the SIA dataset over a wider energy range and parametrised light quarks separately. An update by the same group was given in ref.~\cite{Albino:2008fy}, further
extending the SIA dataset and including $pp$-collision data. The most recent determination was made by the NPC23 group~\cite{Gao:2025bko}, which included in their NLO analysis SIA, SIDIS, and $pp$-collision data. Finally, we note that all extractions mentioned above (DSV, AKK08, NPC23) extracted FFs for the combination $\Lambda+\bar\Lambda$ and thus did not achieve a separation between $\La$ and $\bar\La$. 

In this work, we present the first global analysis at NNLO accuracy in perturbative QCD of the collinear unpolarised FFs of $\La$ hyperons. The fit is based on SIA data, as well as on both neutral- and charged-current SIDIS data. The latter, here considered for the first time, allows us to reliably extract FFs for $\La$ and $\bar\La$ separately.

The paper is organised as follows. In section \ref{sec:Theoretical framework}, we discuss the theoretical framework, with particular emphasis on the newly included charged-current SIDIS cross sections.
In section~\ref{sec:Experimental data} we present the complete dataset of the analysis, with details on the choices made by the experimental collaborations and our treatment. 
The methodological framework is introduced in section~\ref{sec:Fit methodology}. 
Section~\ref{sec:MAPFF_Lambda fit} presents the \maplambda\ fit. We discuss the results of the analysis in section~\ref{subsec:results}, the impact of imposing positivity in section~\ref{subsec:positivity}, and the comparison with other FF sets in section~\ref{subsec:Comparison with other fragmentation function extractions}. Conclusions are outlined in section~\ref{sec:Conclusions}. Appendix~\ref{app:Data-predictions comparison} contains the comparison between data and \maplambda\ predictions for all included datasets. 

\section{Theoretical framework}
\label{sec:Theoretical framework}

In this section, we discuss the theoretical framework on which the analysis is based. The computation of SIA cross sections strictly follows ref.~\cite{Bertone:2017tyb}. For the calculation of the SIDIS cross section, we rely on refs.~\cite{Bonino:2024qbh, Bonino:2025qta}, where NNLO massless QCD corrections in the \(\overline{\text{MS}}\) scheme are computed. The calculations include both electroweak neutral- and charged-current interactions. 

\begin{figure}[t]
\centering
\begin{subfigure}{0.48\textwidth}
    \centering
    \includegraphics[width=0.9\linewidth]{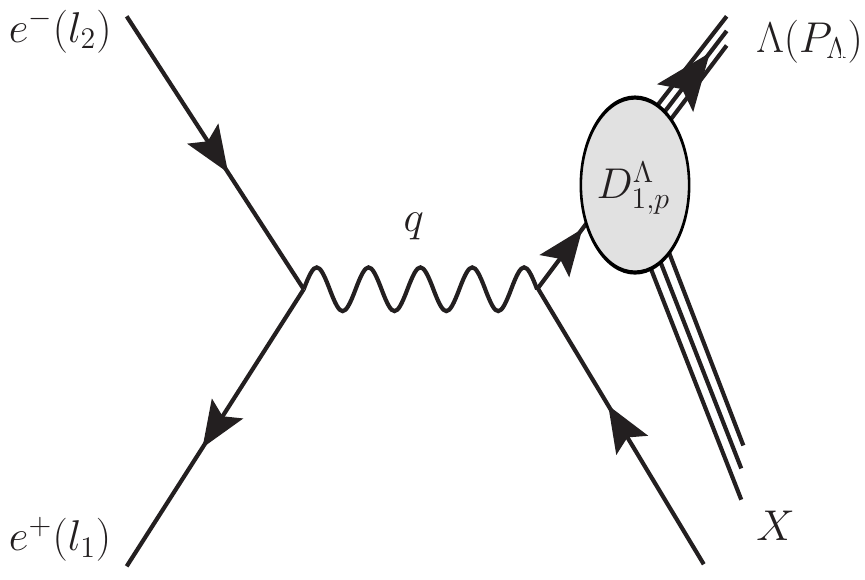}
    \vspace{0.2cm}
    \caption{SIA process with $\La$ production.}
    \label{fig:sia}
\end{subfigure}
\hfill
 \begin{subfigure}{0.5\textwidth}
\centering
    \includegraphics[width=\linewidth]{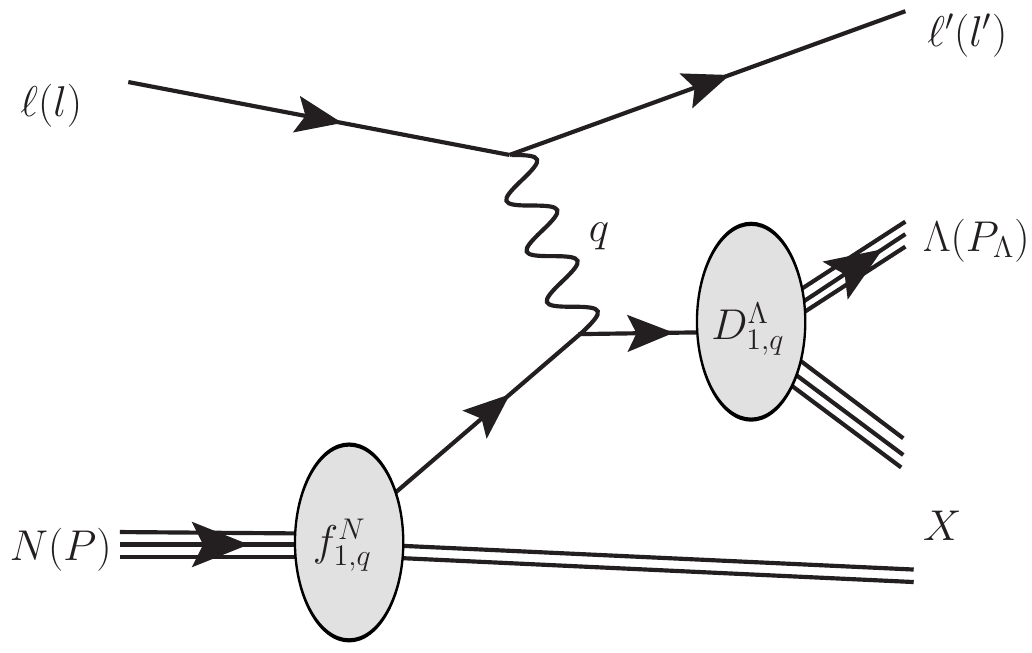}
    \caption{SIDIS process with $\La$ production.}
    \label{fig:sidis}
\end{subfigure}
\caption{A graphical representation of the two processes considered in this analysis: in the left panel single-inclusive electron-positron annihilation (SIA), in the right panel semi-inclusive deep-inelastic lepton-nucleon scattering (SIDIS).}
\end{figure}

\subsection{Single-inclusive annihilation}
\label{subsec:sia}
The single-inclusive production of a $\La$ in electron-positron annihilation, illustrated in figure \ref{fig:sia}, is defined as
\begin{equation}
    e^+(l_1) + e^-(l_2) \rightarrow \La(P_{\La}) + X.
\end{equation}
The relevant Lorentz invariants of this process are 
\begin{equation}
\begin{aligned}
        z\; = \frac{2 P_\La \cdot q}{Q^2},\qquad 
        Q^2 = q^2 > 0, \qquad 
        \sqrt{s}& = Q,
\end{aligned}
\end{equation}
where $z$ can be interpreted as the fraction of energy of the parent parton carried  by the $\La$ and $q=l_1+l_2$ is the momentum of the virtual boson that mediated the process (see figure~\ref{fig:sia}). 

The cross section differential in \(z\) for this process can be written as 
\begin{equation}
    \frac{d\sigma^\La}{dz} = \frac{4\pi\alpha^2}{Q^2} F_2^\La(z,Q),
\end{equation}
where $\alpha(Q)$ is the fine-structure constant. The structure function $F_2^\La(z,Q)$ obeys collinear factorisation, which is valid for $Q \gg \Lambda_{\rm QCD}$ and reads
\begin{equation}
\label{F2_SIA}
    \begin{aligned}
        F^\La_2 (z,Q) =& \left(\frac{1}{n_f} \sum^{n_f}_i \hat{e}^2_i(Q)\right) \left[C^{\rm S}_{2,g} \left(z, \alpha_s (Q)\right) \otimes D^\La_{1,g} \left(z,Q\right)\right. \\
        &\left.+\;C^{\rm S}_{2,q} \left(z, \alpha_s (Q)\right) \otimes D^\La_{1,\Sigma} \left(z,Q\right)+C^{\rm NS}_{2,q} \left(z, \alpha_s (Q)\right) \otimes D^\La_{1,\rm NS} \left(z,Q\right)\right],
    \end{aligned}
\end{equation}
where $\otimes$ denotes the Mellin convolution defined as
\begin{equation}
    C(z) \,\otimes\,D_1(z) = \int_z^1 \frac{dz'}{z'} C(z') D_1\left(\frac{z}{z'}\right).
\end{equation}
In eq.~\eqref{F2_SIA}, $\hat{e}_i$ is the effective electroweak charge of the $i$-th quark flavour (see, e.g., ref.~\cite{deFlorian:1997zj}), $n_f$ is the number of active flavours at the scale $Q$, and $\alpha_s(Q)$ is the strong coupling. The coefficient functions $C^{\rm S}_{2,g}$, $C^{\rm S}_{2,q}$, and $C^{\rm NS}_{2,q} $ are computable in perturbation theory and are convoluted with $D^\La_{1,\Sigma}$, $D^\La_{1,g}$, and $D^\La_{1,\rm NS}$, which are appropriate FF combinations (see ref.~\cite{Bertone:2017tyb} for details). Explicit expressions for the coefficient functions up to NNLO accuracy can be found in refs.~\cite{Rijken:1996vr, Rijken:1996ns, Rijken:1996npa, Blumlein:2006rr, Mitov:2006wy}.

\subsection{Semi-inclusive deep-inelastic scattering}
\label{subsec:sidis}
The semi-inclusive production of a $\La$ in lepton-nucleon deep-inelastic scattering, illustrated in figure \ref{fig:sidis}, is given by the reaction 
\begin{equation}
\ell(l) + N(P) \rightarrow \ell'(l') + \Lambda(P_{\Lambda}) + X.
\end{equation}
The invariants of the process are 
\begin{equation}
    \begin{aligned}
        -Q^2 = q^2 = (l-l')^2, \quad
        x = \frac{Q^2}{2 P\cdot q}, \quad
        z = \frac{P \cdot P_h}{P \cdot q}, \quad
        y = \frac{P \cdot q}{P \cdot l}, \quad
        W^2 = (P+q)^2,
    \end{aligned}
\end{equation}
where $x$ and $z$ can be interpreted as the initial- and final-state momentum fractions, respectively, $y$ is the inelasticity, and $W^2$ is the invariant mass of the target-vector--boson system. These invariants can be related to the collision centre-of-mass energy ${s} = (P+l)^2$ via the relation $Q^2\simeq xys$, which holds when neglecting the target mass. 

On top of the more common neutral-current reaction mediated by a $\gamma^*/Z$ vector boson (see ref.~\cite{Khalek:2021gxf} for a short review), we also consider the charged-current case mediated by $W^\pm$. Indeed, data for the reactions $\nu_{\mu}(\bar\nu_{\mu}) + N \rightarrow \mu^-(\mu^+) + \Lambda + X$ are currently available, which we include in our analysis.

The charged-current SIDIS cross section can be expressed in terms of structure functions as follows:
\begin{equation}
    \frac{d\sigma^{\La,\rm{CC}}}{dx dQ dz} = \frac{4 \pi \alpha^2 }{Q^3 x } 4\eta_W \left[ Y_+ \,{F}_T^{\La}(x,z,Q) + 2 (1-y) \,{F}_L^{\La}(x,z,Q)
    +  l_{\ell} Y_-\,x{F}_3^{\La}(x,z,Q) \right],
\end{equation}
where $l_{\ell}$ is the lepton number. Moreover,  $Y_\pm = 1 \pm (1-y)^2$ and
\begin{equation}
\eta_W =  \frac{1}{2} \left( \frac{G_F M_W^2}{4\pi \alpha}\frac{Q^2}{Q^2 + M_W^2} \right)^2, 
\end{equation}
where $G_F$ is the Fermi constant and $M_W$ the mass of the $W$ boson.
For $Q\gg \La_{\rm QCD}$, structure functions obey collinear factorisation:
\begin{equation}
    F_i^{\Lambda}(x,z,Q) = \sum_{p,p'}\,C_{pp'}(x,z,\alpha_s(Q)) \otimes f_{1,p}^{N}\left(x,Q\right) \otimes D_{1,p'}^{\La}\left(z,Q \right),\quad i=T,L,3,
    \label{eq.SIDIS_SF}
\end{equation}
where the indices $p,{p'}$ run over all active partons at the scale $Q$, $f_{1,p}^{N}$ are the collinear parton distribution functions (PDFs) of the nucleon $N$, and the convolution signs are defined as
\begin{equation}
    C(x,z) \otimes f_1(x)\otimes D_1(z) = \int_x^1 \frac{dx'}{x'} \int_z^1 \frac{dz'}{z'} C(x',z') f_1\left(\frac{x}{x'}\right) D_1\left(\frac{z}{z'}\right).
\end{equation}
The coefficient functions $C_{pp'}$ are computable in perturbative QCD and are presently known up to NNLO in $\alpha_s$. They can be found in refs.~\cite{Bonino:2024qbh, Bonino:2025qta} and are implemented in the \verb|APFEL++| library~\cite{Bertone:2013vaa, Bertone:2017gds} for both neutral- and charged-current structure functions.

\section{Choice of experimental data}
\label{sec:Experimental data}

The {\maplambda} dataset comprises data from SIA and SIDIS. While SIA measurements are delivered as the sum of $\La$ and $\bar\La$ baryons, SIDIS data are mostly provided separately for the two. This distinction allows us to explore for the first time the production of the two baryons individually. The set of included experiments consists of almost all datasets that are currently available and suitable for use, with the exception of H1 2008~\cite{H1:2009pkh} and DELPHI at 183 GeV and 189 GeV~\cite{DELPHI:2000ahn}, which we are unable to describe.\footnote{This was observed for DELPHI also in the other works, where the same choice was made.} 

Measurements for both SIA and SIDIS are often reported as multiplicities, dividing the differential cross section by the total hadronic cross section \(\sigma_h\) or the inclusive DIS cross section, respectively. We compute these denominators using \verb|APFEL++|~\cite{Bertone:2013vaa, Bertone:2017gds}. 

Moreover, measurements are not always provided as differential in \(z\), as presented in section~\ref{sec:Theoretical framework}. In SIA they are often expressed in terms of $x_p=z/ \beta =|\textbf{P}_\Lambda|/(\sqrt{s}/2)$, where $\textbf{P}_\Lambda$ is the three-momentum of the $\La$ in the $e^+e^-$ centre of mass.\footnote{The Belle measurement~\cite{Belle:2017caf} uses the definition $x_p=|\textbf{P}_\Lambda|/\sqrt{s/4-M_{\Lambda}^2}$.} 
Likewise, some SIDIS measurements are differential in $x_F$ or $x_P$. The former variable is defined as $x_F=2P_{\Lambda,L}/W$, where $P_{\Lambda,L}$ is the longitudinal component of the $\Lambda$ momentum in the centre-of-mass frame of the virtual boson and the incoming proton. However, it can also be computed in the frame in which the exchanged virtual boson is purely space-like with 3-momentum $\mathbf{q}=(0,0,-Q)$, commonly denoted as Breit frame. In this case, it will be represented by the latter variable $x_P=2P^{\rm{Breit}}_{\Lambda}/Q$. 

Such alternative variable choice is handled by following the procedure outlined in refs.~\cite{Albino:2008fy, Bertone:2017tyb}, i.e., computing the differential cross sections multiplied by a Jacobian factor that includes corrections arising from finite hadron masses. 
Specifically, for the case of SIDIS, when the cross section is differential in $x_F$, we relate it --- neglecting transverse momentum of the outgoing hadron --- to \(z\) through 
\begin{equation}
\label{eq:z_xf}
    z(x_F) \simeq \frac{x_F}{2}\left(1+\sqrt{1+\frac{4M_{\La}^2}{W^2 x_F^2}}\;\right),
\end{equation}
which corresponds to a leading-order approximation of kinematics~\cite{deFlorian:1997zj}.
Similarly, when the cross section is differential in $x_P$, we have
\begin{equation}
\label{eq:z_xP}
    z(x_P) \simeq \frac{x_P}{2}\left(1+\sqrt{1+\frac{4M_{\La}^2}{Q^2 x_P^2}}\;\right).
\end{equation}
In both cases, keeping in mind that $W^2\equiv W^2(x,Q^2)$, the bin-by-bin computation of $z(x_F)$ and $z(x_P)$ is done at the average values of $Q^2$ and $x$.
\begin{figure}[t]
    \centering
\includegraphics[width=10cm]{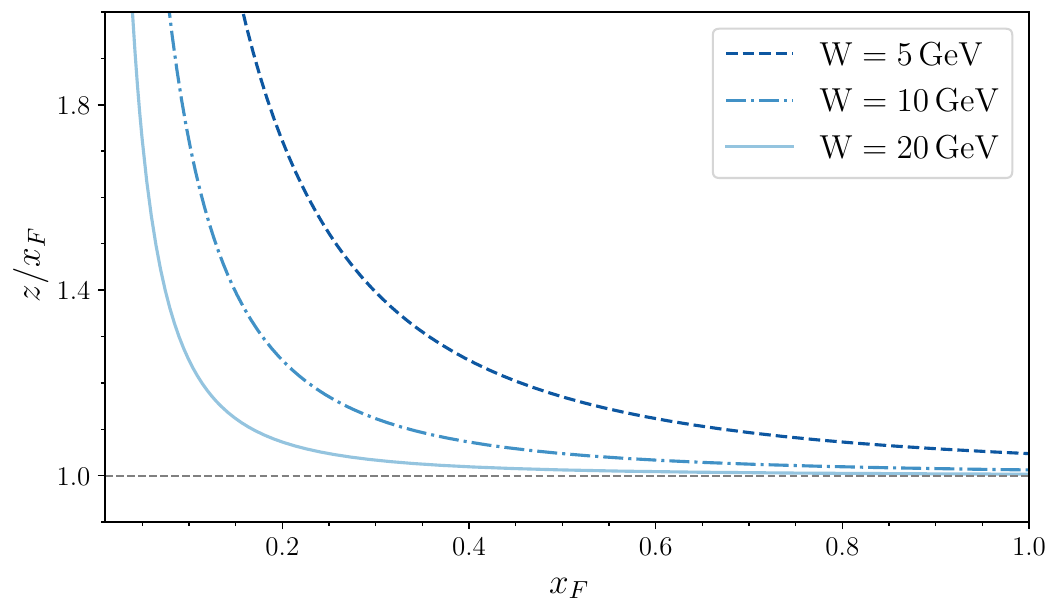}
    \caption{Ratio of $z/x_F$ versus $x_F$ for $W = 5,\,10,\,20$ GeV.}
    \label{fig:xFz}
\end{figure}
Figure \ref{fig:xFz} depicts the ratio of $z$ over $x_F$ as a function of $x_F$ for three different values of $W$. From the plot we can infer that the corrections are sizeable for small values of $x_F$, and hence their inclusion is expected to lead to a better description of the data in this region.
Moreover, while we have verified that the leading-order approximation in eqs.~\eqref{eq:z_xf} and~\eqref{eq:z_xP} is valid at the typical average values of $p_{hT}^2/W^2$ and $p_{hT}^2/Q^2$ specified by the collaborations, this prescription might not be applicable for other SIDIS data sets used in PDF or FF extractions.

The variable $x_F$ allows for a more careful study of $\La$ production, as it provides information on the region in which the particle is produced. In particular, restricting the analysis to positive values of $x_F$ favours hadrons produced in the \emph{current fragmentation} region, i.e., the fragmentation of the struck quark (cf.~figure \ref{fig:sidis}), in contrast to the \emph{target fragmentation} region, where hadrons emerge from the remnants of the target nucleon. Only in the first case the interpretation of the process in terms of PDFs and single-hadron FFs outlined in section~\ref{subsec:sidis} applies.

We now move to discussing the datasets included in the present analysis. As for SIA, we include results from CERN (ALEPH~\cite{ALEPH:1996oqp}, DELPHI at 91.2 GeV~\cite{DELPHI:1993vpj}, OPAL~\cite{OPAL:1996gsw}), from DESY (TASSO at 14, 22, and 34 GeV~\cite{TASSO:1984nda}, 33.3 GeV~\cite{TASSO:1981uqa}, 34.8 and 42.1 GeV~\cite{TASSO:1988qlu}, ARGUS~\cite{ARGUS:1987ogo}, CELLO~\cite{CELLO:1989adw}), from SLAC (MARK II~\cite{delaVaissiere:1984xg}, HRS~\cite{Baringer:1986jd}, SLD~\cite{SLD:1998coh}), and from KEK (Belle~\cite{Belle:2017caf}). 
Table \ref{table:sia} lists all datasets along with additional information: the corresponding reference, the measured observable, the center-of-mass energy $\sqrt{s}$, and the number of data points that pass the selection criteria. We observe that SIA data spans a range in $\sqrt{s}$ from 10 GeV up to $M_Z$ always combining \(\La\) and \(\bar{\La}\).

\begin{table}
\footnotesize
    \setlength{\tabcolsep}{10pt} 
    \centering
    \begin{tabular}{c c c c c}
    \hline
    \noalign{\vskip 1ex}
    \multirow{1}{*}{Experiment}
    & \multirow{1}{*}{ Ref. } & \multirow{1}{*}{Observable} & \multirow{1}{*}{$\sqrt{s}$ [GeV]} & \multirow{1}{*}{$N_{dat}(\La+\bar\La)$}  \\
    \noalign{\vskip 1ex}
    \hline 
    \noalign{\vskip 1ex}
    ARGUS &~\cite{ARGUS:1987ogo}  & $\frac{1}{\beta\sigma_{h}} \frac{d\sigma}{dz}$ & 10.0  & 16 \\ [1ex] 
     Belle &~\cite{Belle:2017caf} &  $\frac{d\sigma}{dx_p} \mathrm{\scriptscriptstyle[nb]}$& 10.52   &  15 \\ [1ex] 
    TASSO14 &~\cite{TASSO:1984nda} &  $\frac{s}{\beta} \frac{d\sigma}{dz}\mathrm{\scriptscriptstyle[\mu b]}$ & 14.0 & 3 \\[1ex] 
    TASSO22 &~\cite{TASSO:1984nda} & $\frac{s}{\beta} \frac{d\sigma}{dz}\mathrm{\scriptscriptstyle[\mu b]}$ & 22.0 & 4 \\[1ex] 
    TASSO33.3 &~\cite{TASSO:1981uqa}  &  $\frac{d\sigma}{dp}\mathrm{\scriptscriptstyle[pb]}$& 33.3 &   5 \\ [1ex] 
    TASSO34 &~\cite{TASSO:1984nda} &  $\frac{s}{\beta} \frac{d\sigma}{dz}\mathrm{\scriptscriptstyle[\mu b]}$& 34.0 & 7 \\[1ex] 
    TASSO34.8 &~\cite{TASSO:1988qlu} &  $\frac{s}{\beta} \frac{d\sigma}{dz}\mathrm{\scriptscriptstyle[nb]}$& 34.8  & 10 \\[1ex] 
    TASSO42.1 &~\cite{TASSO:1988qlu} &  $\frac{s}{\beta} \frac{d\sigma}{dz}\mathrm{\scriptscriptstyle[nb]}$& 42.1  & 4 \\[1ex] 
    HRS &~\cite{Baringer:1986jd} &  $\frac{s}{\beta} \frac{d\sigma}{dz}\mathrm{\scriptscriptstyle[nb]}$& 29.0   & 12 \\ [1ex]
    MARK II &~\cite{delaVaissiere:1984xg} &  $\frac{s}{\beta} \frac{d\sigma}{dz}\mathrm{\scriptscriptstyle[nb]}$& 29.0 & 13 \\[1ex] 
    CELLO &~\cite{CELLO:1989adw} &  $\frac{1}{\beta \sigma_{h}} \frac{d\sigma}{dz}$& 35.0 & 7 \\[1ex] 
    ALEPH &~\cite{ALEPH:1996oqp} &  $\frac{1}{\sigma_{h}} \frac{d\sigma}{dx_p}$& 91.2 & 25 \\[1ex] 
    DELPHI91 &~\cite{DELPHI:1993vpj} &  $\frac{1}{\sigma_{h}} \frac{d\sigma}{dx_p}$& 91.2 &  10 \\[1ex] 
    OPAL &  ~\cite{OPAL:1996gsw} &  $\frac{1}{\sigma_{h}} \frac{d\sigma}{dz}$& 91.2 &  15 \\ [1ex]
    SLD & ~\cite{SLD:1998coh}  &  $\frac{1}{\sigma_{h}} \frac{d\sigma}{dx_p}$& 91.2 &  15 \\ [1ex] 
     & ~\cite{SLD:1998coh}  &  $\frac{1}{\sigma_{h}} \frac{d\sigma}{dx_p}{|_{uds}}$& 91.2 & 8 \\ [1ex] 
     & ~\cite{SLD:1998coh}  &  $\frac{1}{\sigma_{h}} \frac{d\sigma}{dx_p}{|_{c}}$& 91.2 &  8 \\ [1ex] 
     & ~\cite{SLD:1998coh}  &  $\frac{1}{\sigma_{h}} \frac{d\sigma}{dx_p}{|_{b}}$& 91.2 &  8 \\ [2ex]
    \hline 
    \noalign{\vskip 1ex}
     Total & & & & 185\\
    \noalign{\vskip 1ex}
    \hline 
    \end{tabular}
    \caption{SIA datasets included in the analysis. We list the experiments and their observables. We also report the centre-of-mass energy and the number of data points per experiment after applying the selection criteria.
  \label{table:sia}}
  \end{table}

Some additional comments are in order. The Belle Collaboration provides differential inclusive cross sections with and without initial-state QED radiative corrections. We found that the inclusion of radiative corrections leads to a better description.
Concerning the SLD experiment, tagged and untagged datasets are available. They are classified as inclusive hadronic $Z$ decays (untagged), as well as separately for $Z$ decays into primary light (up, down, strange), charm, and bottom flavours.  

For the first time in $\Lambda$ FF analyses, SIDIS measurements for both neutral- and charged-current interactions are included. For the neutral-current case, we use data from H1~\cite{H1:1996kfw} and ZEUS~\cite{ZEUS:2011cdi} at DESY, from the Chicago, Harvard, Illinois, Oxford (CHIO) Collaboration~\cite{Hicks:1980cj} and E665~\cite{ E665:1993dzd} at Fermilab, as well as from EMC~\cite{EuropeanMuon:1984rkv} at CERN.
Regarding the charged-current SIDIS data, included datasets are from the {Aachen-Bonn-CERN-Munich-Oxford (ABCMO) Collaboration}~\cite{Aachen-Bonn-CERN-Munich-Oxford:1981mcl}, WA59~\cite{WA59:1991cop}, and NOMAD~\cite{NOMAD:2001kdr}, all hosted at CERN. 

The SIDIS datasets included in this analysis, along with the respective kinematic details, are listed in table \ref{table:sidis}.
Observables are generally expressed in terms of the number $N$ of particles or events. Since $N$ is related to the cross section through the luminosity ($N=\sigma \cdot \mathcal{L}$) and differential distributions $dN/dz$ are reported as divided by $N_{\rm DIS}$, the observables listed in table \ref{table:sidis} can be interpreted as multiplicities. NOMAD data make an exception because differential cross sections are given in ``arbitrary units".  Indeed, for this dataset distributions are normalised in a way that they integrate to unity. For this reason, we normalise predictions accordingly and only fit the shape of the distributions.

When computing predictions, we integrate both numerator and denominator of the multiplicity separately over the accepted phase space, also taking into account our selection criteria. 
For E665, EMC, and the charged-current SIDIS datasets, we assume $Q^2_{\rm{min}} \le Q^2 \le s y_{\rm max}$, since no $Q_{\rm{max}}$ is provided. For the CHIO measurements and for all of the charged-current SIDIS datasets, an $x$ range is not given. Therefore, we estimate it from the kinematic coverage of the experiments, including the constraints imposed by the range in inelasticity $y$ and the minimum value of the invariant mass $W$.

\begin{table}[t!]
\footnotesize
    \centering
    \begin{tabular}{c c c c c c}
    \hline
    \noalign{\vskip 1ex}
    \multirow{1}{*}{Experiment}
    & \multirow{1}{*}{ Ref. } & \multirow{1}{*}{Observable} & \multirow{1}{*}{$\sqrt{s}$ [GeV]} & \multirow{1}{*}{Constraints} & \multirow{1}{*}{$N_{dat}(\La)+N_{dat}(\bar\La)$}  \\
    \noalign{\vskip 1ex}
    \hline 
    \noalign{\vskip 1ex}
     Neutral-Current SIDIS & & & & \\[0.6ex]
    \hline 
    \noalign{\vskip 1ex}
    CHIO &~\cite{Hicks:1980cj} & $ \frac{1}{N_{\rm DIS}}\frac{dN_{\La,\bar{\La}}}{dz}$ & 20.6 &  $0.08 < y < 0.93$ & 3 + 4\\[1ex]
    EMC &~\cite{EuropeanMuon:1984rkv}  & $\frac{1}{N_{\rm DIS}} \frac{dN_{\La,\bar{\La}}}{dx_F}$ &  23.5 & $\begin{array}{c} 
    0.07 < y < 0.9\\
    W>4~\text{GeV} 
    \end{array}$ & 4 + 3 \\[2.5ex]
    E665 &~\cite{ E665:1993dzd} & $\frac{1}{N_{\rm DIS}} \frac{dN_{\La,\bar{\La}}}{dx_F}$ & 30.3 & 
    $\begin{array}{c} 
    0.1 < y < 0.85 \\
    W>10~\text{GeV}
    \end{array}$ & 3 + 3 \\[2.5ex]
    H1 1996 &~\cite{H1:1996kfw}   & $ \frac{1}{N_{\rm DIS}}\frac{dN_{\La+\bar{\La}}}{dx_F}$ & 300 & $0.05 < y < 0.6$  & 2\\[1ex]
    ZEUS $10\!-\!40$ &~\cite{ZEUS:2011cdi}  & $ \frac{1}{N_{\rm DIS}}\frac{dN_{\La+\bar{\La}}}{dx_P}$ & 319 &  $0.04 < y < 0.95$ & 1\\[1ex]
    ZEUS $40\!-\!160$ &~\cite{ZEUS:2011cdi}  & $ \frac{1}{N_{\rm DIS}}\frac{dN_{\La+\bar{\La}}}{dx_P}$ & 319 &  $0.04 < y < 0.95$ & 2\\[1ex]
    ZEUS $160\!-\!640$ &~\cite{ZEUS:2011cdi}  & $ \frac{1}{N_{\rm DIS}}\frac{dN_{\La+\bar{\La}}}{dx_P}$ & 319 &  $0.04 < y < 0.95$ & 4\\[1ex]
    ZEUS $640\!-\!2560$ &~\cite{ZEUS:2011cdi}  & $ \frac{1}{N_{\rm DIS}}\frac{dN_{\La+\bar{\La}}}{dx_P}$ & 319 &  $0.04 < y < 0.95$  & 3\\[1ex]
    ZEUS $2560\!-\!10240$&~\cite{ZEUS:2011cdi}  & $ \frac{1}{N_{\rm DIS}}\frac{dN_{\La+\bar{\La}}}{dx_P}$ & 319 &  $0.04 < y < 0.95$ & 1\\ [1ex]
    \hline
    \noalign{\vskip 1ex}
    Charged-Current SIDIS & & & & \\[0.6ex]
    \hline 
    \noalign{\vskip 1ex}
    ABCMO &~\cite{Aachen-Bonn-CERN-Munich-Oxford:1981mcl}  & $\frac{1}{N_{\rm DIS}} \frac{dN_{\La}}{dz}$ &  9.0 & $\begin{array}{c} 
    y < 0.93\\
    W>1.5~\text{GeV} 
    \end{array}$  & 4 + 0\\[2.5ex]
    WA59 &~\cite{WA59:1991cop} & $ \frac{1}{N_{\rm DIS}}\frac{dN_{\La}}{dx_F}$ & 9.2 &  
    $\begin{array}{c} 
    y < 0.89\\
    W>\sqrt{5}~\text{GeV} 
    \end{array}$ & 6 + 0\\[2.5ex]
    NOMAD &~\cite{NOMAD:2001kdr}& $\frac{1}{N_{\La,\bar{\La}}} \frac{dN_{\La,\bar{\La}}}{dz}$ & 9.3 & 
    $\begin{array}{c} 
    y < 0.89\\
    W>\sqrt{2}~\text{GeV}
    \end{array}$  & 6 + 7 \\[2.5ex]
    \hline 
    \noalign{\vskip 1ex}
     Total & & & & & 56\\
    \noalign{\vskip 1ex}
    \hline 
    \end{tabular}
    \caption{Neutral- and charged-current SIDIS datasets included in the analysis. We list the experiments, the observables, the centre-of-mass energy, limits on \(y\) and \(W\) imposed by the experiments, and the number of data points per experiment after applying the selection criteria. The last reports two values when both data for $\Lambda$ and $\bar\Lambda$ are available, including the case of only $\La$ data. On the other hand, when only one value is reported, data are relative to the sum $\Lambda+\bar\Lambda$. }
    \label{table:sidis}
    \end{table}

Our selection criteria are as follows. In order to guarantee the applicability of perturbation theory, we require $Q^2\ge1 \text{ GeV}^2$ for all datasets. As far as SIA data are concerned, we also require $z_{\rm min}<z<0.9$, with $z_{\rm min}=0.075$ for all experiments with $\sqrt{s}< M_Z$, and $z_{\rm min}=0.02$ for those with $\sqrt{s}= M_Z$. Exceptionally, we set $z_{\rm min}=0.1$ for MARK II and $z_{\rm min}=0.03$ for DELPHI because the description of the data is very poor in the excluded regions. 

\begin{figure}[ht]
    \centering
    \includegraphics[width=10cm]{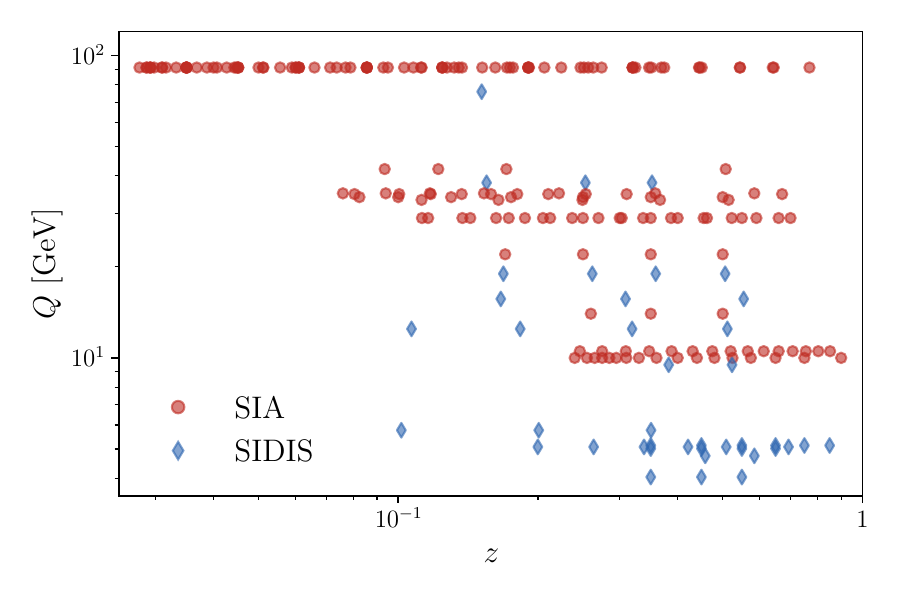}
    \caption{Kinematic coverage in the $(z,Q)$ plane of the included datasets. Red circles depict SIA data, while
    light blue diamonds label the included SIDIS data points, which are plotted at the average $Q$ for each bin.}
    \label{image:kincov}
\end{figure}

For SIDIS data, we impose $z_{\rm min}=0.1$ and $z_{\rm max}=0.9$ in the case of cross sections differential in $x_F$ or $x_P$, as these variables allow for the best description and separation between current and target fragmentation regions. 
By contrast, for cross sections differential in $z$, we impose more restrictive constraints. Moreover, we adopt two different strategies for $\Lambda$ and $\bar\Lambda$ data: $\Lambda$ production is significantly more favoured in the target fragmentation region, since the remnants of the target can contribute to two of the three required valence quarks for the $\Lambda$ to form. 
On the other hand, for $\bar\Lambda$ the two regions are usually populated in a more symmetric way.  
For these reasons, for $\Lambda$ cross sections differential in $z$, it is more likely than for the $\bar\Lambda$ to get sizeable contributions from target fragmentation at low values of $z$. We thus chose $z_{\rm min}=0.3$ for $\Lambda$ data versus $z_{\rm min}=0.2$ for $\bar\Lambda$ data. Exceptions are made for the ZEUS bins at low $Q^2$, specifically for $10 \text{ GeV}^2<Q^2<40\text{ GeV}^2$ and $40 \text{ GeV}^2<Q^2<160\text{ GeV}^2$, we set the allowed range to $0.4<z<0.8$ and $0.3<z<0.8$, respectively. For $160 \text{ GeV}^2<Q^2<640\text{ GeV}^2$, we set $0.1<z<0.8$. 
All these kinematic choices are summarised in table \ref{table:cuts}. The kinematic coverage of the resulting dataset included in the fit is shown in figure~\ref{image:kincov}. For SIDIS, plotted values correspond to the average $Q$ for each bin.

\begin{table}
\footnotesize
\begin{center}
    \begin{tabular}{c c c c}
    \hline
    \noalign{\vskip 1ex}
    Process &  & $Q$ range & $z$ range \\
    \noalign{\vskip 1ex}
    \hline
    \noalign{\vskip 1ex}
    \multirow{2}{*}{SIA} & $\sqrt{s}< M_Z$  & \multirow{2}{*}{$Q=\sqrt{s}$} & $0.075<z<0.9$ \\[0.3ex]
     & $\sqrt{s}= M_Z$ & & $0.02<z<0.9$ \\[1ex]
    \hline
     \noalign{\vskip 1ex}
    \multirow{3}{*}{SIDIS} & $\Lambda$ , $\frac{d\sigma}{dz}$&  & $0.3<z<0.9$\\[0.3ex] 
     & $\bar\Lambda$, $\frac{d\sigma}{dz}$ & $Q \ge 1 \text{ GeV}$ & $0.2<z<0.9$\\[0.3ex] 
     & $\frac{d\sigma}{dx_F}$ &  & $0.1<z<0.9$\\[1ex] 
     \hline
    \end{tabular}
    \end{center}
    \caption{Kinematic constraints imposed for SIA and SIDIS data. For the former, we separate the processes for values of $\sqrt{s}$. For the latter, we rely on the natural separation of the regions given by $x_F$ to impose a less restrictive constraint, while for cross sections differential in $z$ we exclude a larger region at low $z$.}
    \label{table:cuts}
  \end{table}

The CHIO dataset includes an uncertainty whose nature is not specified and which we treat as uncorrelated.
Both the TASSO (at 34.8 GeV and at 42.1 GeV) and the ZEUS datasets list experimental values with asymmetric uncertainties. We treat these data following the procedure defined in ref.~\cite{DAgostini:2003syq}. 

\section{Fit methodology}
\label{sec:Fit methodology}

The methodological framework used to extract FFs from data follows that of refs.~\cite{Bertone:2017tyb, Khalek:2021gxf, AbdulKhalek:2022laj}, which in turn builds on the methodology originally developed by the NNPDF Collaboration~\cite{NNPDF:2021njg}. It rests on three main pillars: the propagation of experimental uncertainties into FFs through Monte Carlo sampling of the experimental data, the parametrisation of FFs by means of a neural network, and the optimisation of the parameters using backpropagation and gradient descent minimisation.

Concerning the Monte Carlo sampling, all of our FFs sets are made of $N_{\rm rep} = 100$ replicas. This choice ensures that the Monte Carlo ensemble faithfully captures the main statistical properties of the experimental data, i.e., central values, uncertainties, and correlations estimated as averages, standard deviations, and covariances, respectively, over the replica sample.

As for the parametrisation, we use a one-layered feed-forward neural network $\mathcal{N}(z,\bm{\theta})$, where $\bm{\theta}$ denotes the set of free parameters, with one input node corresponding to the momentum fraction $z$, ten nodes in the hidden layer, and seven output nodes corresponding to the independent distributions (see below). We use a sigmoid activation function for all layers but the output, the latter being linear. This architecture amounts to a total of 97 free parameters.

The inclusion of both neutral- and charged-current SIDIS data --- the latter for the first time for a $\La$ FF analysis --- allows for a more accurate flavour separation that SIA data alone do not provide.
Furthermore, data on \(\bar{\Lambda}\), through charge conjugation, provide further constraints on flavour-antiflavour separation.
This allows us to treat the valence quarks $u$, $d$, and $s$ for the first time as independent, which motivates our choice of fitting the following seven distributions:
\begin{equation}
     D_{1,u}^{\Lambda}, \quad D_{1,d}^{\Lambda},\quad  D_{1,s}^{\Lambda}, \quad D_{1,\bar{u}}^{\Lambda}=D_{1,\bar{d}}^{\Lambda}=D_{1,\bar{s}}^{\Lambda},\quad  D_{1,c^+}^{\Lambda},\quad  D_{1,b^+}^{\Lambda}, \quad  D_{1,g}^{\Lambda},
     \label{eq:setofFFs}
 \end{equation}
where the notation $q^+ = q + \bar{q}$ is adopted. The parametrisation scale is set to $\mu_0=5$~GeV.

The kinematic constraint $D_{1,i}^{\Lambda}\rightarrow 0$ as $z\rightarrow 1$ is enforced by construction through subtraction of the neural network evaluated at $z=1$:
\begin{equation}
    zD_{1,i}^{\Lambda} (z, \mu_0) = \mathcal{N}_i\,( z, \bm{\theta}) -\mathcal{N}_i\,( 1, \bm{\theta} ),
    \label{eq:ffpar}
\end{equation}
where $i$ labels the independent distributions listed in eq.~(\ref{eq:setofFFs}). While eq.~\eqref{eq:ffpar} is our default choice, an alternative parametrisation is also considered, namely
\begin{equation}
        zD_{1,i}^{\Lambda} (z, \mu_0) = \left[ \mathcal{N}_i\,( z, \bm{\theta}) -\mathcal{N}_i\,( 1, \bm{\theta} ) \right]^2.
    \label{eq:pospar}
\end{equation}
The purpose of squaring the r.h.s. is to enforce positivity at the initial scale. 
Its impact on the fit will be discussed in section~\ref{subsec:positivity}.

Parameter optimisation is carried out following the same strategy as in refs.~\cite{Khalek:2021gxf,AbdulKhalek:2022laj}. For each replica, we minimise the $\chi^2$ (whose definition is given in ref.~\cite{Khalek:2021gxf}) using the Levenberg--Marquardt algorithm as implemented in \texttt{Ceres-Solver}~\cite{Agarwal_Ceres_Solver_2022}, computing
the gradient of the neural network with respect to the free parameters analytically using the {\sc NNAD} library~\cite{AbdulKhalek:2020uza}. To avoid overfitting, we adopt the cross-validation strategy with a training fraction of 50\%. Replicas that, at the end of the training, result in a value of the $\chi^2$
per data point larger than three are discarded.\footnote{In practice, out of the 110 replicas usually generated for each extraction always less than ten are discarded.}

Finally, as a PDF set used to compute the SIDIS cross sections during the fits we chose \texttt{NNPDF31\_(n)nlo\_pch\_as\_0118}~\cite{NNPDF:2017mvq}, with perturbative accuracy consistently matched to that of the FF determination. Heavy-quark mass effects are treated within the zero-mass variable-flavour-number scheme (ZM-VFNS)~\cite{Thorne:2000zd}.

\section{The {\maplambda} fit}
\label{sec:MAPFF_Lambda fit}
In this section, we present  the {\maplambda} baseline fit, discuss an alternative parametrisation, and compare our results to previously published $\Lambda$ FFs.
\subsection{Results}
\label{subsec:results}

The fit quality of both NLO and NNLO extractions is reported in table \ref{table:chi2}. 
We list the experimental datasets along with the number of data points that pass the selection criteria and the corresponding $\chi^2$ values per data point ($\chi^2/N_{\rm dat}$). 
We obtain a global $\chi^2/N_{\rm dat}$ of 1.09 for a total of 241 data points. 
The majority of the data come from SIA experiments, where $\Lambda$ production was studied more extensively. 
The datasets from TASSO at $34.8$ GeV and from DELPHI show the poorest agreement between data and fits. 
In the case of SIDIS, the dataset from the ABCMO Collaboration is the most poorly described.
By contrast, the NOMAD measurement provides the most statistically significant contribution, with the highest number of data points amongst SIDIS measurements.

\begin{table}[t]
\footnotesize
\setlength{\tabcolsep}{16pt} 
\centering
\begin{tabular}{c c c c}
\hline\noalign{\vskip 1ex}
\multirow{2}{*}{Experiment} &
\multirow{2}{*}{$N_{\rm dat}$ } & $\chi^2/N_{\rm dat}$ & $\chi^2/N_{\rm dat}$\\
& & NLO & NNLO \\
\noalign{\vskip 1ex}
\hline
Neutral-Current SIDIS  \\
\hline
CHIO  $\Lambda$ & 3 & 0.56 & 0.55 \\[0.3ex] 
CHIO  $\bar{\Lambda}$ & 4 & 1.11 & 1.03 \\[0.3ex] 
EMC $\Lambda$ & 4 & 1.19 & 0.83\\[0.3ex]  
EMC $\bar{\Lambda}$ & 3 & 0.12 & 0.47\\[0.3ex]  
E665 $\Lambda$ & 3 & 0.50 & 0.48\\[0.3ex]  
E665 $\bar{\Lambda}$ & 3 & 1.00 & 1.27\\[0.3ex] 
H1 $\Lambda \bar{\Lambda}$ & 2 & 1.18 & 1.01\\[0.3ex]  
ZEUS $\Lambda \bar{\Lambda}$ 10-40  & 1 & $<\, 0.01$ & $<\, 0.01$ \\[0.3ex] 
ZEUS $\Lambda \bar{\Lambda}$ 40-160 & 2 & 0.61 & 0.75\\[0.3ex] 
ZEUS $\Lambda \bar{\Lambda}$ 160-640 & 4 & 1.07 & 0.81 \\[0.3ex] 
ZEUS $\Lambda \bar{\Lambda}$ 640-2560 & 3 & 0.46 & 0.37 \\[0.3ex] 
ZEUS $\Lambda \bar{\Lambda}$ 2560-10240 & 1 & 0.03 & 0.03 \\[0.3ex]
\hline
Charged-Current SIDIS \\ 
\hline
WA59 $\Lambda$ & 6 & 0.34 & 0.33 \\  [0.3ex]
NOMAD $\Lambda$ & 6 & 1.36 & 1.85 \\[0.3ex]  
NOMAD $\bar{\Lambda}$ & 7 & 0.14 & 0.18\\[0.3ex]  
ABCMO $\Lambda$ & 4 & 4.05 & 3.05 \\  [0.3ex]
\hline
SIA\\ 
\hline
ARGUS & 16 & 1.49 & 1.55 \\ [0.3ex]
Belle & 15 & 1.13 & 1.26 \\ [0.3ex]
TASSO 14 & 3 & 0.23 & 0.23 \\ [0.3ex]
TASSO 22 & 4 & 0.48 & 0.49 \\ [0.3ex]
TASSO 33 & 5 & 0.78 & 0.78 \\ [0.3ex]
TASSO 34 & 7 & 0.57 & 0.56 \\ [0.3ex]
TASSO 34.8 & 10 & 2.79 & 2.75 \\ [0.3ex]
TASSO 42.1 & 4 & 1.13 & 1.12 \\ [0.3ex]
HRS & 12 & 0.82 & 0.82 \\ [0.3ex]
MARK II & 13 & 1.47 & 1.47 \\ [0.3ex]
CELLO & 7 & 0.52 & 0.51 \\ [0.3ex]
ALEPH & 25 & 0.48 & 0.44 \\[0.3ex]
DELPHI 91.2 & 10 & 2.62 & 2.58 \\ [0.3ex]
OPAL & 15 & 1.04 & 1.04 \\ [0.3ex]
SLD & 15 & 0.24 & 0.25 \\ [0.3ex]
SLD UDS & 8 & 2.30 & 2.02 \\ [0.3ex]
SLD C & 8 & 2.31 & 2.19 \\ [0.3ex]
SLD B & 8 & 0.50 & 0.80 \\ [0.3ex]

\hline\noalign{\vskip 1ex}
Total & 241 & 1.10 & 1.09 \\ \noalign{\vskip 1ex}
\hline\noalign{\vskip 1ex}
\end{tabular}
\caption{List of datasets, their number of included data points after applying the kinematic constraints, as well as their $\chi^2$ per data point at NLO and at NNLO.
\label{table:chi2}}
\end{table}

\begin{figure}[t]
    \includegraphics[width=15cm]{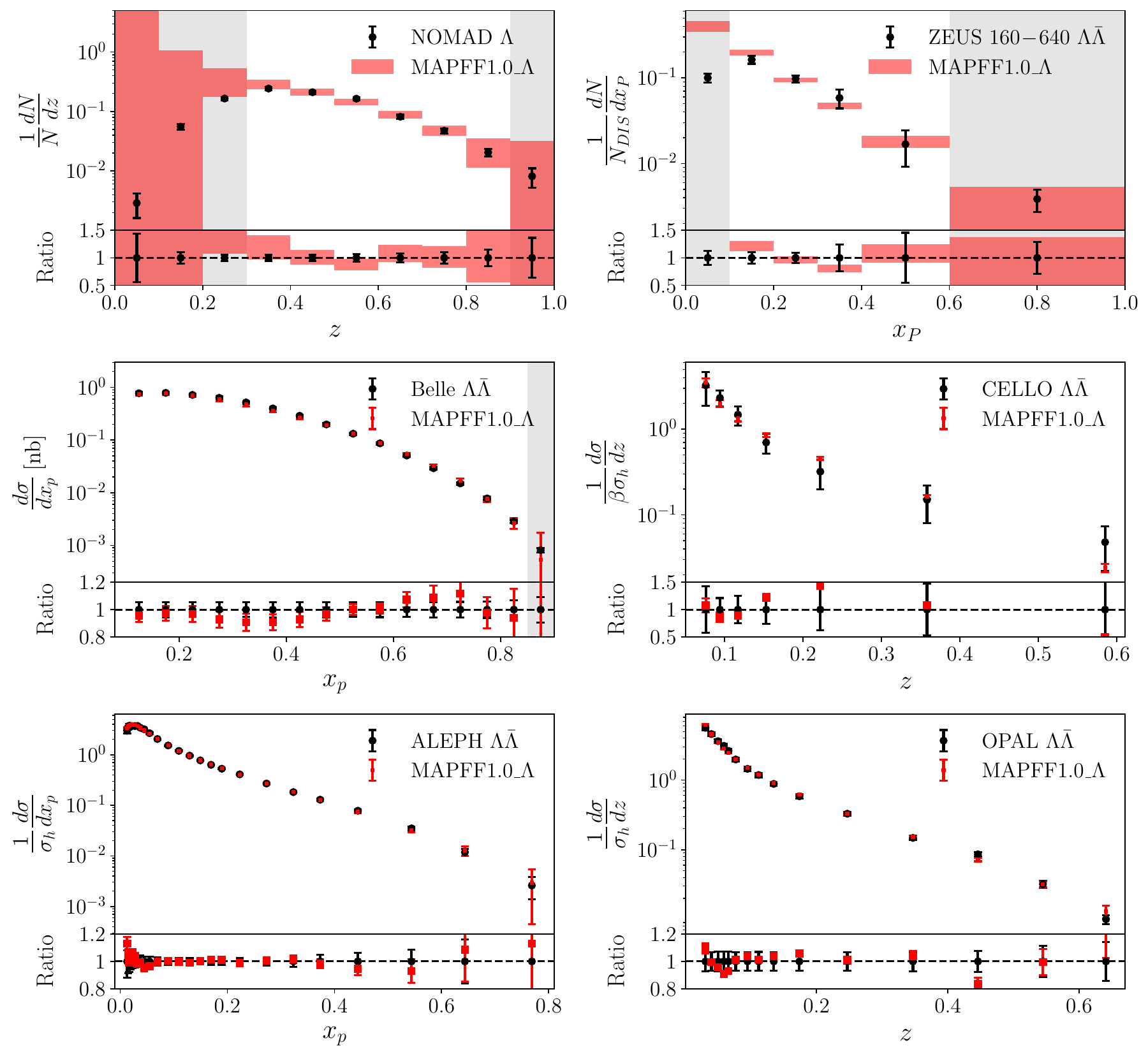}
    \caption{Comparison of selected datasets with the \maplambda\ predictions. In the first row, SIDIS data are shown: charged-current NOMAD  data to the left and neutral-current ZEUS data to the right. The second and third rows show comparisons of SIA data, at $\sqrt{s}=10.52$ GeV for Belle, $\sqrt{s}=35$ GeV for CELLO, and $\sqrt{s}=M_Z$ for ALEPH and OPAL. The lower panels display the corresponding ratios of the \maplambda\ values to the measured ones. Shaded areas in grey at low and/or high values of $z$ represent the regions excluded from the fit due to the imposed kinematic constraints.
   When the observable of the relative dataset is given for a bin in $z$, the prediction is represented by a band, otherwise it is computed at the average $z$ value given by the experiment. 
    \label{image:datavstheory}}
\end{figure}

While the complete set of comparisons plots between the experimental data and predictions is presented in appendix~\ref{app:Data-predictions comparison}, here 
we comment on a selection of datasets, shown in figure~\ref{image:datavstheory}.
Each plot displays the comparison between \maplambda\ predictions and data, also presented as ratios to data in the lower panels. The shaded areas represent the regions excluded from the fit due to the selection criteria discussed in section \ref{sec:Experimental data}.

When the observable of the relative dataset is given in bins of $z$, predictions are represented as bands, otherwise they are displayed as points.  
The upper row of figure~\ref{image:datavstheory} shows the charged-current SIDIS dataset from NOMAD (left) and neutral-current SIDIS dataset from ZEUS (right).
The second and third rows are instead devoted to SIA data, with the second row displaying Belle at $\sqrt{s}=10.52$ GeV (left) and CELLO at $\sqrt{s}=35$ GeV (right), and the third row displaying  ALEPH and OPAL, which are both at  $\sqrt{s}=M_Z$. In all cases we observe a good description of the data.

\begin{figure}
    \centering
    \includegraphics[width=0.85\linewidth]{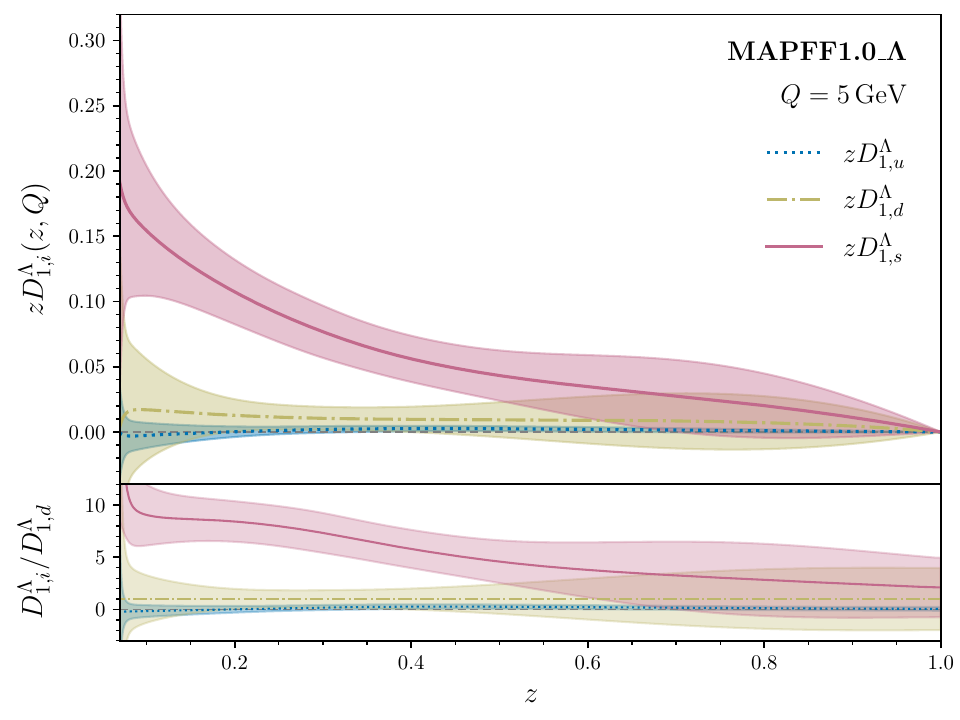}
    \caption{Comparison between the FFs for the $\La$ valence quarks, i.e., $D_{1,i}^{\Lambda}$ ($i=u,d,s$), at $Q=5 \text{ GeV}$, at NNLO in perturbative corrections. The dominance of the $s$ quark is observed with respect to the $u$ and $d$ quarks.}
    \label{fig:FFs}
\end{figure}

In figure \ref{fig:FFs}, we show the comparison between the FFs for $u$, $d$, and $s$, plotting central value and standard deviation. We observe that the strange quark is dominant over the entire $z$ range. This behaviour is consistent with the valence quark composition of the $\Lambda$: having a strange quark as a valence component, either the fragmenting parton is already a strange quark or strangeness has to be acquired in the hadronisation process. Due to the significantly heavier mass of the $s$ quark compared to both $u$ and $d$ quarks, the latter mechanism is known to be suppressed (``strangeness suppression"). As such, fragmentation of $u$ or $d$ quarks into $\Lambda$ hyperons is less pronounced compared to $s$-quark fragmentation.

\begin{figure}[t]
\centering
    \includegraphics[width=\linewidth]{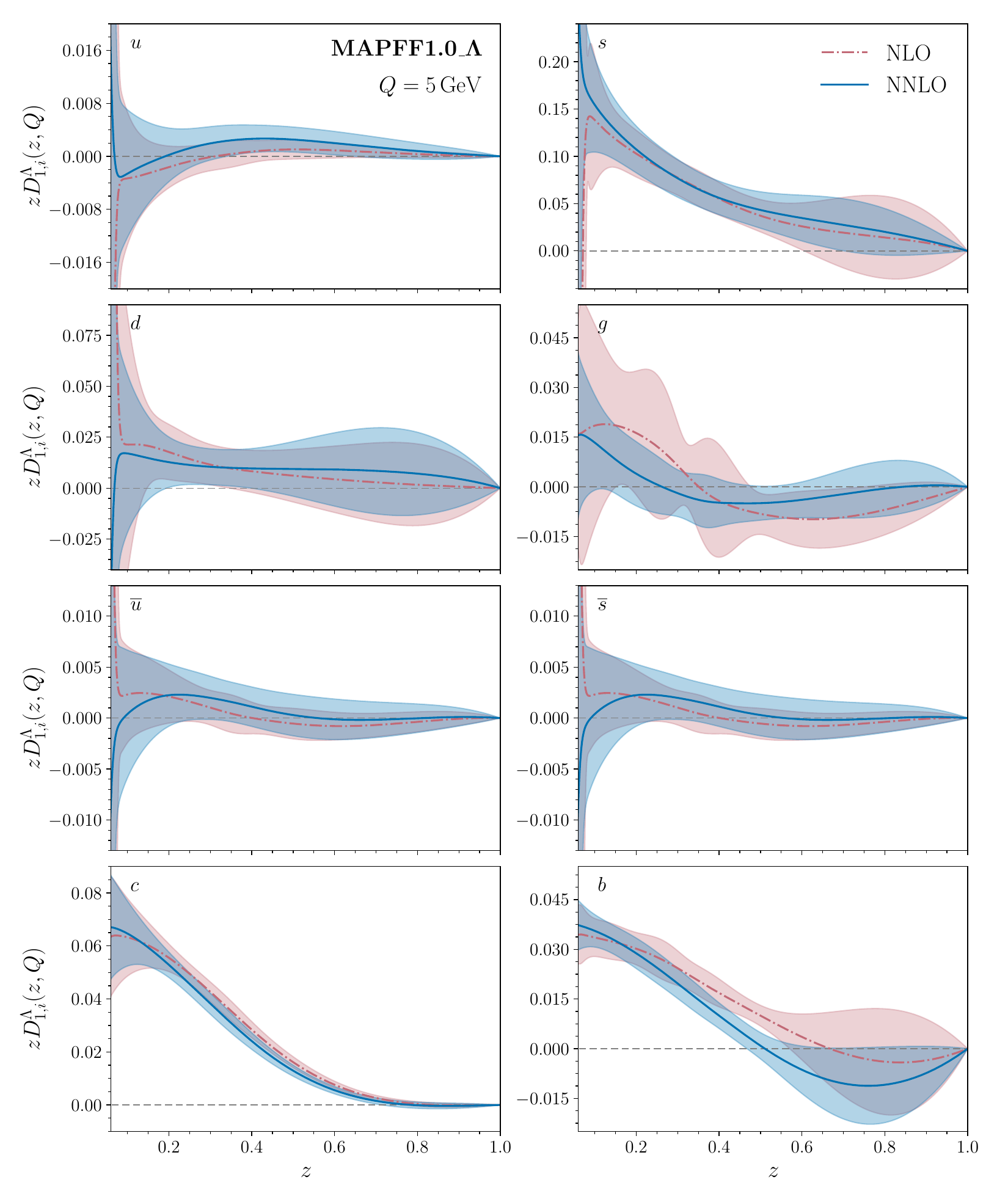}
    \caption{Comparison of the \maplambda\ set at NLO and NNLO, for $u$, $s$, $d$, $g$, $\bar u$, $\bar s (=\bar{u})$, $c$, and $b$. Our FF for $\bar d $ is equivalent to $\bar u$ and $\bar s$, and is not shown.}
    \label{image:NLOvsNNLO}
\end{figure}

Table \ref{table:chi2} also allows us to compare NLO and NNLO fits. The quality of the fit remains stable with increasing perturbative accuracy, with a slight improvement in the description of the majority of the datasets observed at NNLO.
In figure~\ref{image:NLOvsNNLO}, we show the comparison of the resulting NLO and NNLO FFs for all distributions, except for $D_{1,\bar d}^{\Lambda}$ since it is equal to both $D_{1,\bar u}^{\Lambda}$ and $D_{1,\bar s}^{\Lambda}$. The FFs for the light quarks exhibit a slight positive shift as the perturbative accuracy increases. As expected, the gluon FF is not well constrained, and is mostly compatible with zero at the parametrisation scale at both orders. Charm and bottom FFs tend to shift downwards at NNLO. In particular, the bottom FF tends to become negative for large values of $z$, particularly at NNLO, but it also features a large uncertainty, smaller at NNLO, that makes it compatible with zero in this region. A reduction of the uncertainties at high $z$ going from NLO to NNLO is also observed for the strange quark.

\subsection{Imposing positivity}
\label{subsec:positivity}

In section~\ref{sec:Fit methodology} we have introduced in eq.~\eqref{eq:ffpar} the baseline FF parametrisation, along with an alternative parametrisation in eq.~\eqref{eq:pospar}. The latter enforces FF positivity at the initial scale $\mu_0=5$ GeV, which allows us to study the effects of this constraint on FFs.

\begin{figure}
\centering
    \includegraphics[width=\linewidth]{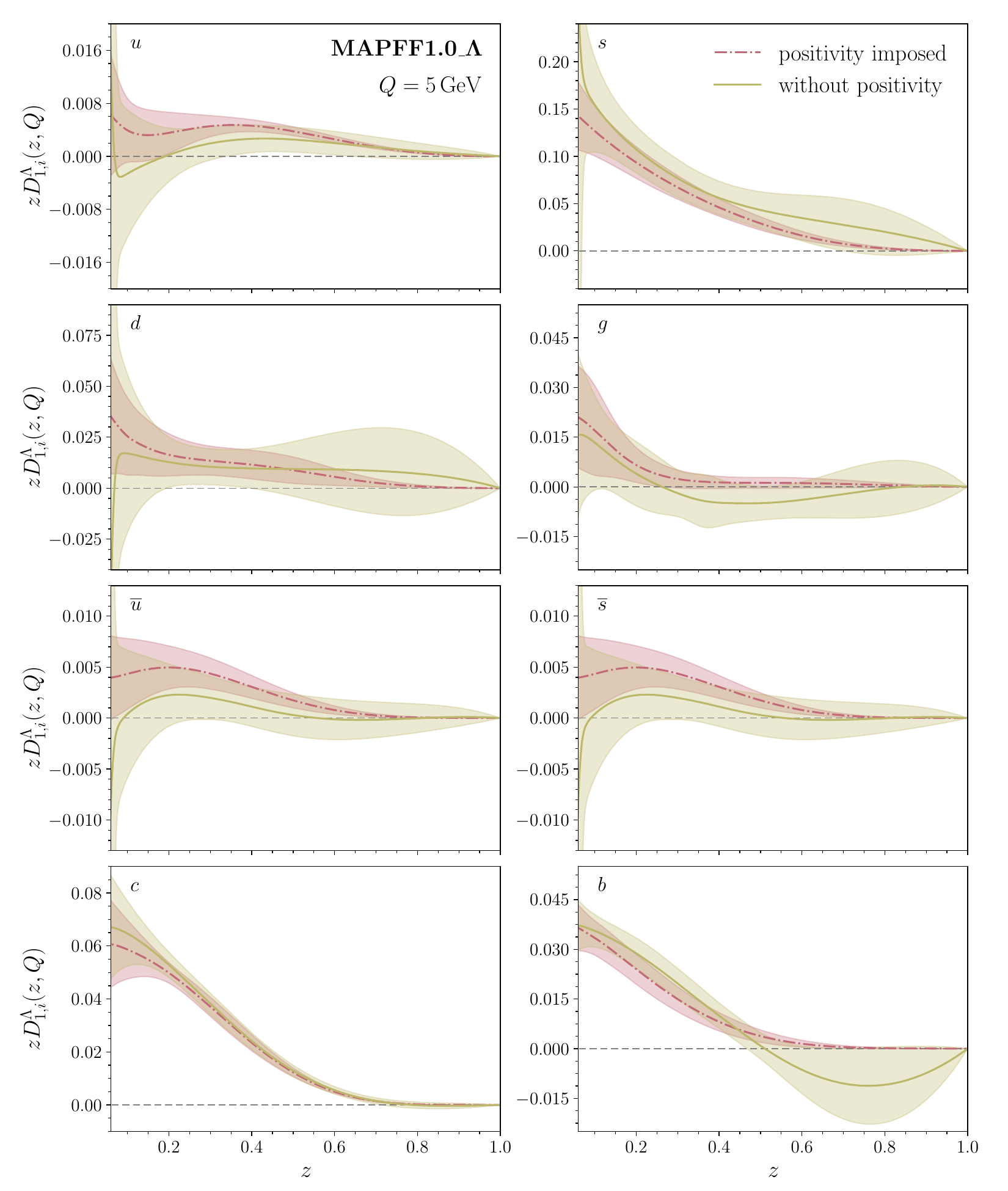}
    \caption{Comparison of the baseline parametrisation of the {\maplambda} FFs at NNLO with an alternative one, which requires the FFs to be positive at the initial scale of  $\mu_0=5$ GeV. The parametrisation comparison is shown for $u$, $s$, $d$, $g$, $\bar u$, $\bar s (=\bar{u})$, $c$, and $b$. The FF for $\bar d $ is equivalent to the FF for $\bar u$ and $\bar s$, and is not shown.}
    \label{image:positivity}
\end{figure}

The fit at NNLO obtained imposing positivity results in a $\chi^2$ of 1.16, i.e., larger than that without positivity. Figure~\ref{image:positivity} shows the comparison between the two parametrisations. 
The main effect of imposing positivity is a reduction in size of the uncertainty over the entire range in $z$. 
This is particularly pronounced for $z \gtrsim 0.5$ and, in the case of light quarks, also for $z<0.1$, where only SIA data provide experimental constraints.
The positivity requirement overall preserves the shape of most of FFs. A more noticeable difference is seen when the central value of the FF is close to zero or negative (for the unconstrained set), which applies to the $u$ quark,  the light anti-quarks, and the $b$ quark at large $z$. 
Imposing positivity then results in a more constrained shape for the replicas and hence for the central values, leading to an upward shift that is compensated by downward shifts for the FFs of other flavours.

\begin{figure}
\centering
    \includegraphics[width=\linewidth]{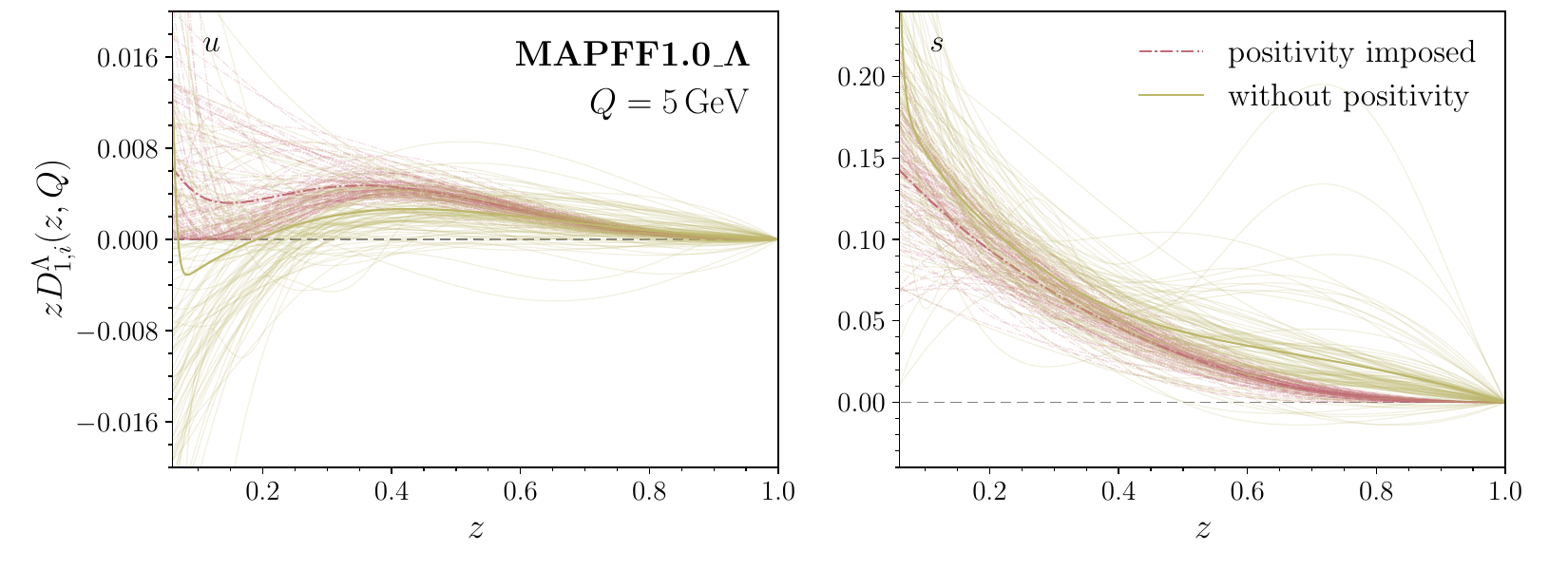}
    \caption{Comparison of the baseline parametrisation of the {\maplambda} FFs at NNLO with an alternative one, which requires the FFs to be positive at the initial scale of $\mu_0=5$ GeV. The parametrisation comparison is done for the complete set of replicas and shown for $u$ and $s$.}
    \label{image:positivity_reps}
\end{figure}

It is instructive to study the effect of positivity by displaying in figure~\ref{image:positivity_reps} the single replicas for up and strange FFs. In the left panel, the distribution of replicas for the up quark at low $z$ has a tendency to become negative for the unconstrained fit, which is not allowed by construction for those with positivity. In the right panel, despite the wider spread of the strange quark replicas without positivity at large $z$, the distributions of replicas in the two cases are compatible. 

\subsection{Comparison with other fragmentation function extractions}
\label{subsec:Comparison with other fragmentation function extractions}
Here, we compare the NLO FFs extracted in this work to those from DSV~\cite{deFlorian:1997zj}, AKK08~\cite{Albino:2008fy}, and NPC23~\cite{Gao:2025bko} at the scale $Q=5$ GeV. 
Unlike our extraction, where for the first time all three valence flavours are treated independently and $D_{1,q}^{\Lambda}$ is directly extracted, the previous ones provide FF sets for only the combination of $\Lambda$ and $\bar\Lambda$, which also imposes $D_{1,q}^{\Lambda+\bar\Lambda}=D_{1,\bar q}^{\Lambda+\bar\Lambda}$. Note that both DSV and AKK08 simply sum the two contributions, i.e., $D_{1,q}^{\Lambda+\bar\Lambda} \equiv D_{1,q}^{\Lambda}+D_{1,q}^{\bar\Lambda}$, while NPC23 uses the average.\footnote{Resulting differences in the FFs from the factor 2 difference in the definitions are taken into account in the comparisons.} In addition, DSV assumes all light quarks and antiquarks to be equal, i.e., $u=\bar u =d=\bar d =s=\bar s$, while NPC23 only requires isospin symmetry, i.e., $u=d$.

\begin{figure}
\centering
    \includegraphics[width=\linewidth]{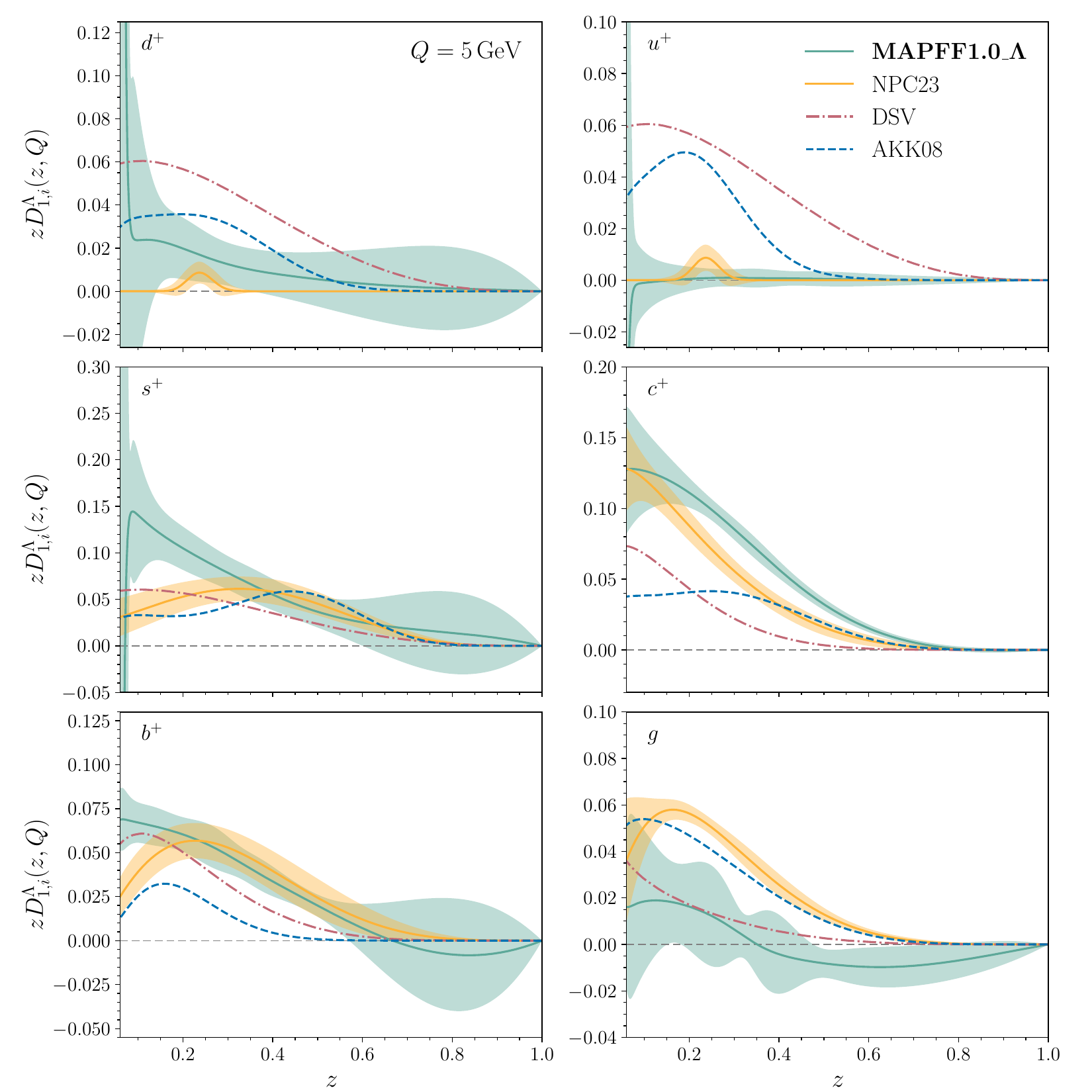}
    \caption{Comparison at NLO of the {\maplambda} FFs with DSV~\cite{deFlorian:1997zj}, AKK08~\cite{Albino:2008fy}, as well as NPC23~\cite{Gao:2025bko}, for $Q=5 \text{ GeV}$ and for $d^+,u^+,s^+,c^+,b^+, g$.}
    \label{image:FF_comparison_5}
\end{figure}
 Besides differences in the choices of independent flavours, all extractions differ in the included datasets. All of them use SIA data but with some differences. Different datasets for $pp$ collisions are included in the AKK08 and NPC23 extractions. The latter also includes SIDIS data from ZEUS. Conversely, the \maplambda\ extraction includes, for the first time, neutral-current and charged-current SIDIS data.

As a consequence of charge-conjugation symmetry,
\begin{equation}
    D_{1,q^+}^{\Lambda}\equiv D_{1,q}^{\Lambda}+D_{1,\bar q}^{\Lambda} = D_{1,q}^{\Lambda+\bar\Lambda}.
\end{equation}
 In figure \ref{image:FF_comparison_5}, we compare our FFs for $D_{1,q^+}^\Lambda$ and $D_{1,g}^\Lambda$ to those from previous extractions. The FF sets from DSV and AKK08 do not provide any uncertainty, which makes a rigorous assessment of the degree of agreement more complicated. Nevertheless, several observations can be made. In particular, we observe a good overall agreement for the FF of the down quark, $D_{1,d^+}^\La$. By contrast, the FF of the up quark, $D_{1,u^+}^\La$, is comparatively suppressed in our analysis. 
 Unlike the previous extractions we do not impose positivity,
while imposing increases the up contribution as seen in section~\ref{subsec:positivity}. As for the strange quark, $D_{1,s^+}^\La$, which we parametrise independently, we observe a dominance over the up and down quarks even without separating $s$ and $\bar s$. This behaviour is consistent with the valence-quark composition of the $\Lambda$, being the strange a valence component in this baryon, in conjunction with the expected strangeness suppression in the fragmentation.

Charm and bottom FFs show a good level of agreement across all four sets.  Finally, the gluon FF $D_{1,g}^\La$ remains poorly constrained in our analysis. Including data from hadron collision should better constrain the FF.
\section{Summary}
\label{sec:Conclusions}

This work presents
a global extraction of $\Lambda$ FFs from experimental data carried out, for the first time, at NNLO accuracy in perturbative QCD. Our analysis includes data for single-inclusive electron-positron annihilation (SIA) and semi-inclusive deep-inelastic scattering (SIDIS). The latter process features, for the first time, both neutral- and charged-current data. This allows for the first determination of $\Lambda$ FFs that treats $\Lambda$ and $\bar\Lambda$ separately, offering new insights into the hadronisation mechanism of strange baryons, and establishing a baseline for future phenomenological and experimental investigations. In particular, our results are suitable for phenomenological applications in the context of TMD extractions.

Section~\ref{sec:Experimental data} summarises the complete dataset of the analysis. To ensure an accurate description of the data across the whole kinematic range, we include hadron-mass corrections, which have a significant impact especially at low values of $z$. In addition, we impose kinematic constraints on $Q^2$ and $z$ aimed at guaranteeing the applicability of our computational framework. In SIDIS, we exploit the ability of the variable $x_F$ to separate current-fragmentation from target-fragmentation region to impose less restrictive kinematic constraints of data differential in this variable. For $z$-differential data, instead, we adopt different strategies for $\La$ and $\bar\La$ data, motivated by the different production rates of the two particles. 

The fit methodology is discussed in section~\ref{sec:Fit methodology}. A statistical framework is adopted based on the Monte Carlo sampling method generating a set of $N_{\rm rep}=100$ data replicas. FFs are parametrised 
using a single one-layered feed-forward neural network.  

In section \ref{sec:MAPFF_Lambda fit}, fit quality at NLO and NNLO is discussed. Based on the analysis of the $\chi^2$, we observe a general stability when increasing the perturbative accuracy, with a slight improvement in the description of the majority of the datasets at NNLO. 
We obtain for the first time FFs for the $\La$ and $\bar\Lambda$ baryon separately, parametrising up, down, and strange quarks FFs independently. We find a dominant behaviour of the strange quark with respect to the other light flavours. Moreover, up and down quark FFs exhibit similar size and shape.

We also consider an alternative parametrisation that imposes positivity on FFs at the parametrisation scale. The main effect consists in a reduction of the uncertainty bands of all the FFs. Expectedly, the shape of FFs is more constrained by positivity when their central values are close to zero, especially in the large-$z$ region.

We also compare our FFs with those from previous extractions. We observe a good overall agreement for the FF of $d+\bar d$, while the $u+\bar u$ combination is more suppressed in our analysis. Thanks to our ability of treating the strange quark FF as an independent distribution, we are able to reveal its characteristic dominance over up and down quarks. 
Alongside, charm and bottom FFs show a good level of agreement across all four sets, while the gluon FF remains poorly constrained.

All results presented in this paper are obtained using the code available at 
\begin{center}\href{https://github.com/MapCollaboration/MontBlanc/tree/Lambda}{https://github.com/MapCollaboration/MontBlanc/tree/Lambda}.
\end{center}
The extracted FFs are made available in the LHAPDF format as \texttt{MAPFF10NLOLambda} and \texttt{MAPFF10NNLOLambda}.  

\acknowledgments
We thank E.R.~Nocera for useful discussions in the early stage of this work. We are grateful to W.~Vogelsang, J.~Gao and C.~Liu, as well as  M.~Zaccheddu for providing us with the published FF sets. 
This work was supported by 
the Basque Foundation for Science (IKERBASQUE), 
the grants IT1977-26 funded by the Basque Government, PID2022-136510NB-C33 and CNS2022-135186 funded by MCIN/AEI/10.13039/501100011033, FSE+ and European Union ``NextGenerationEU''/PRTR, 
and l’Agence Nationale de la Recherche (ANR), project ANR-24-CE31-7061-01,
as well as by the European Union COST action No.~24159 (SHARP).

\appendix
\section{Complete set of data compared to the \maplambda\ fit}
\label{app:Data-predictions comparison}

\begin{figure}[ht]
\centering
\vspace{0.8cm}
    \includegraphics[width=\linewidth]{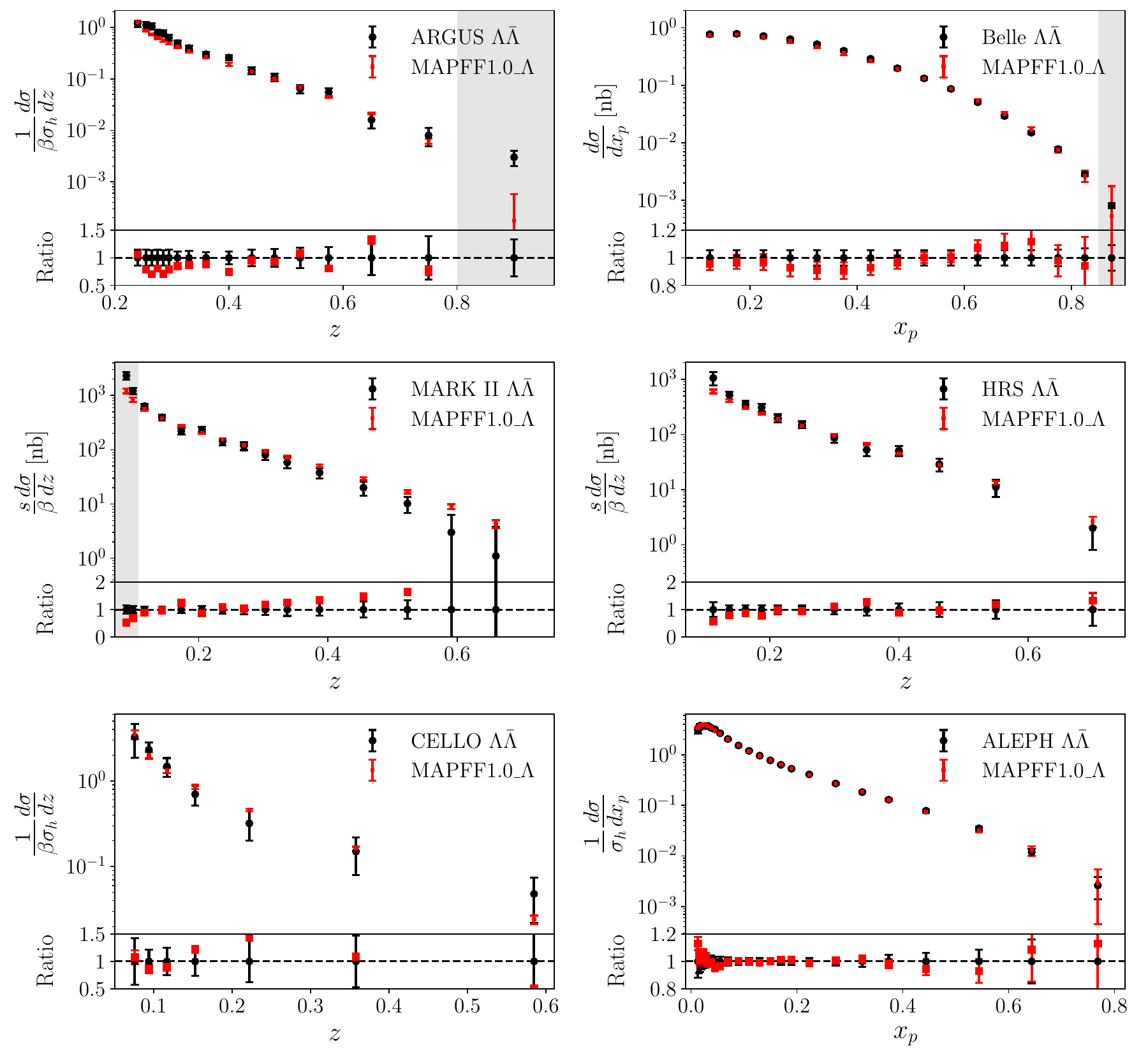}
    \caption{Comparison of selected SIA datasets with the \maplambda\ calculation. In the upper panels, the actual experiment-specific observable is plotted, while the lower panels display the corresponding ratio of the \maplambda\ values to the measured ones. Shaded areas in grey at low or high values of $z$ represent the regions excluded from the fit due to the imposed kinematic constraints. When the observable of the relative dataset is given for a bin in $z$, the prediction is represented by a band, otherwise it is computed at the average $z$ value given by the experiment.}
    \label{image:sia_a}
\end{figure}

\begin{figure}[t]
\centering
    \includegraphics[width=\linewidth]{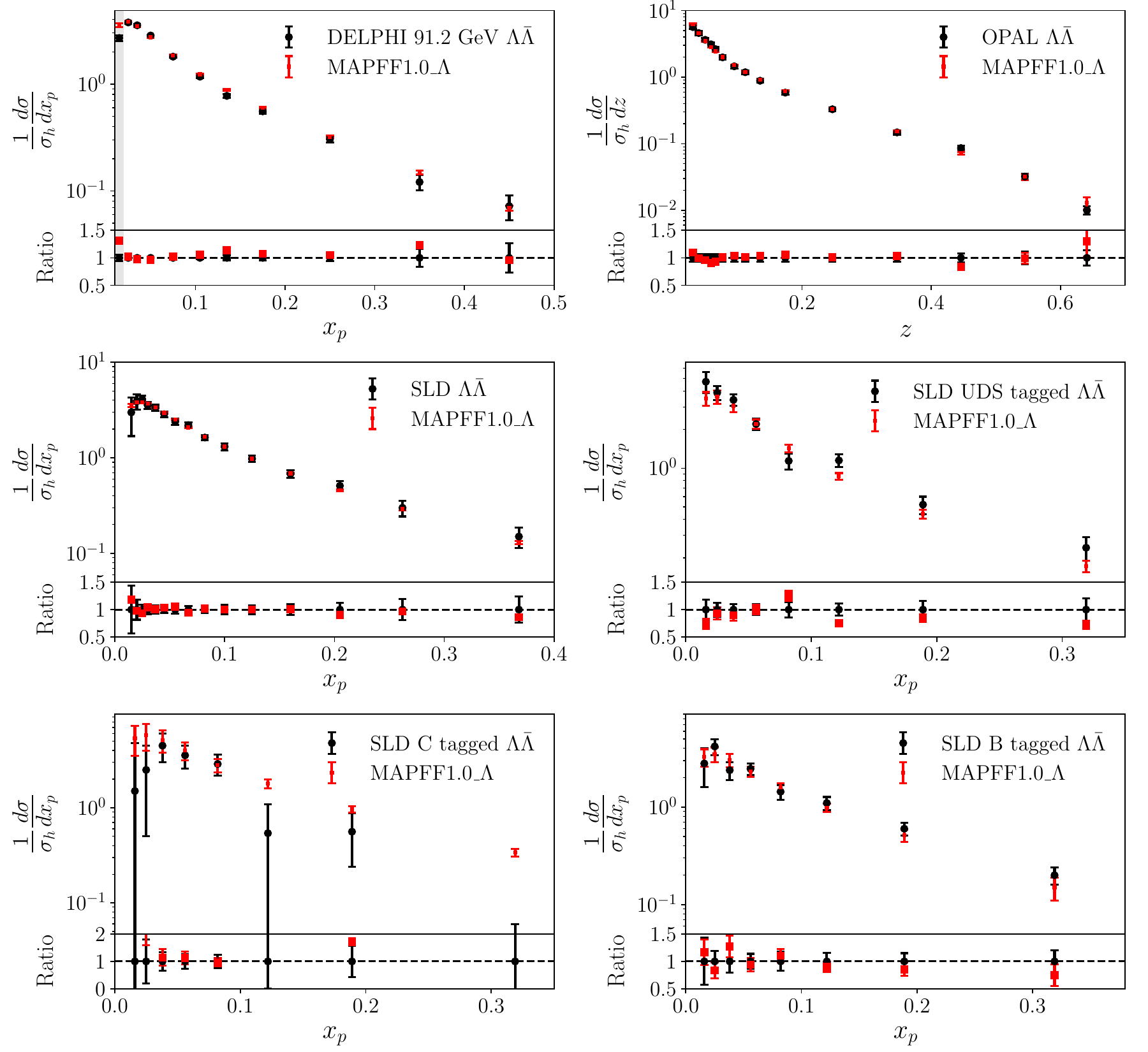}
    \caption{Comparison of selected SIA datasets with the \maplambda\ calculation. In the upper panels, the actual experiment-specific observable is plotted, while the lower panels display the corresponding ratio of the \maplambda\  values to the measured ones. Shaded areas in grey at low or high values of $z$ represent the regions excluded from the fit due to the imposed kinematic constraints. When the observable of the relative dataset is given for a bin in $z$, the prediction is represented by a band, otherwise it is computed at the average $z$ value given by the experiment. }
    \label{image:sia_b}
\end{figure}

\begin{figure}
\centering
    \includegraphics[width=\linewidth]{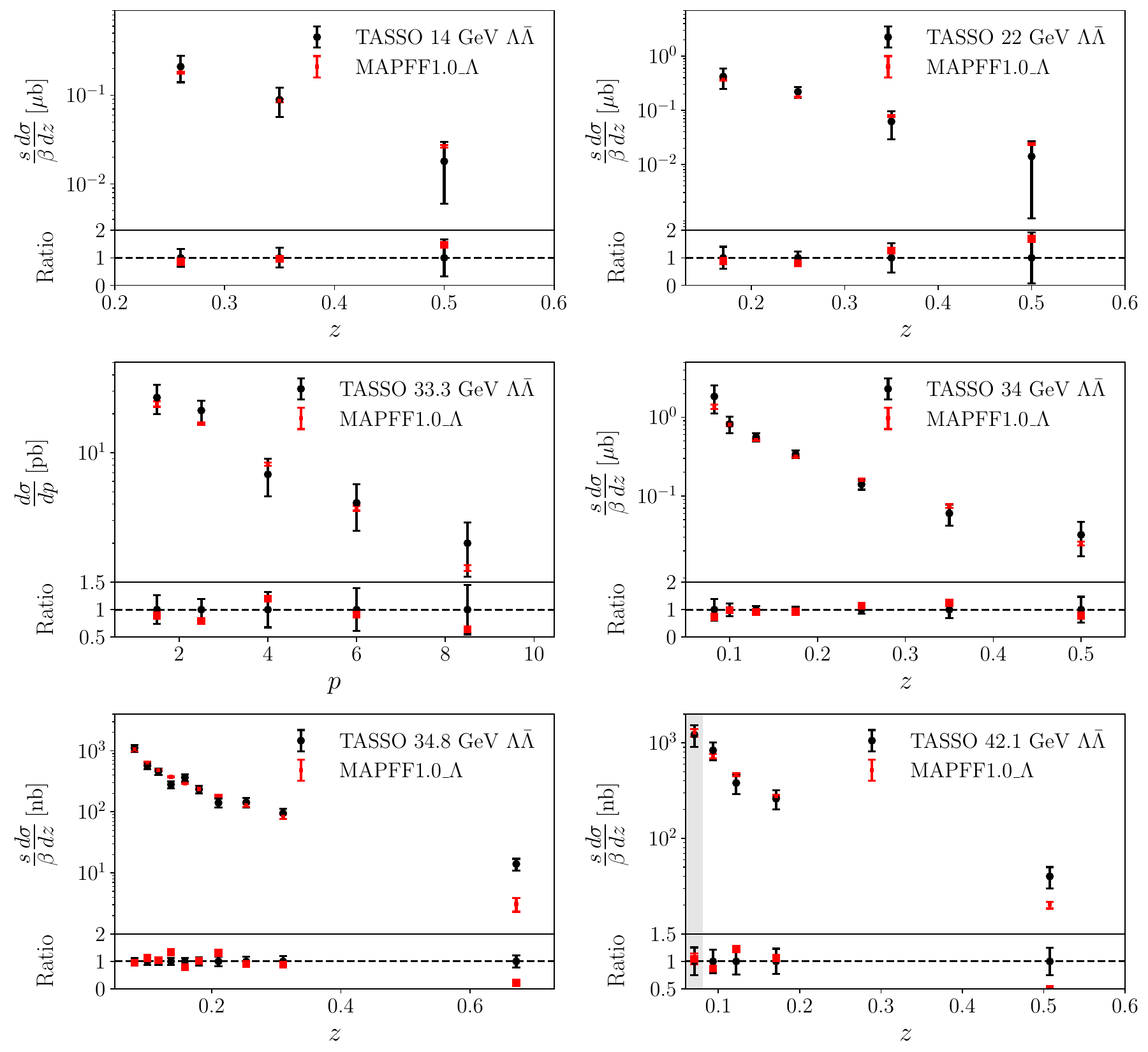}
    \caption{Comparison of selected SIA datasets with the \maplambda\ calculation. In the upper panels, the actual experiment-specific observable is plotted, while the lower panels display the corresponding ratio of the \maplambda\  values to the measured ones. Shaded areas in grey at low or high values of $z$ represent the regions excluded from the fit due to the imposed kinematic constraints. When the observable of the relative dataset is given for a bin in $z$, the prediction is represented by a band, otherwise it is computed at the average $z$ value given by the experiment.  }
    \label{image:sia_c}
\end{figure}

\begin{figure}
\centering
    \includegraphics[width=0.93\linewidth]{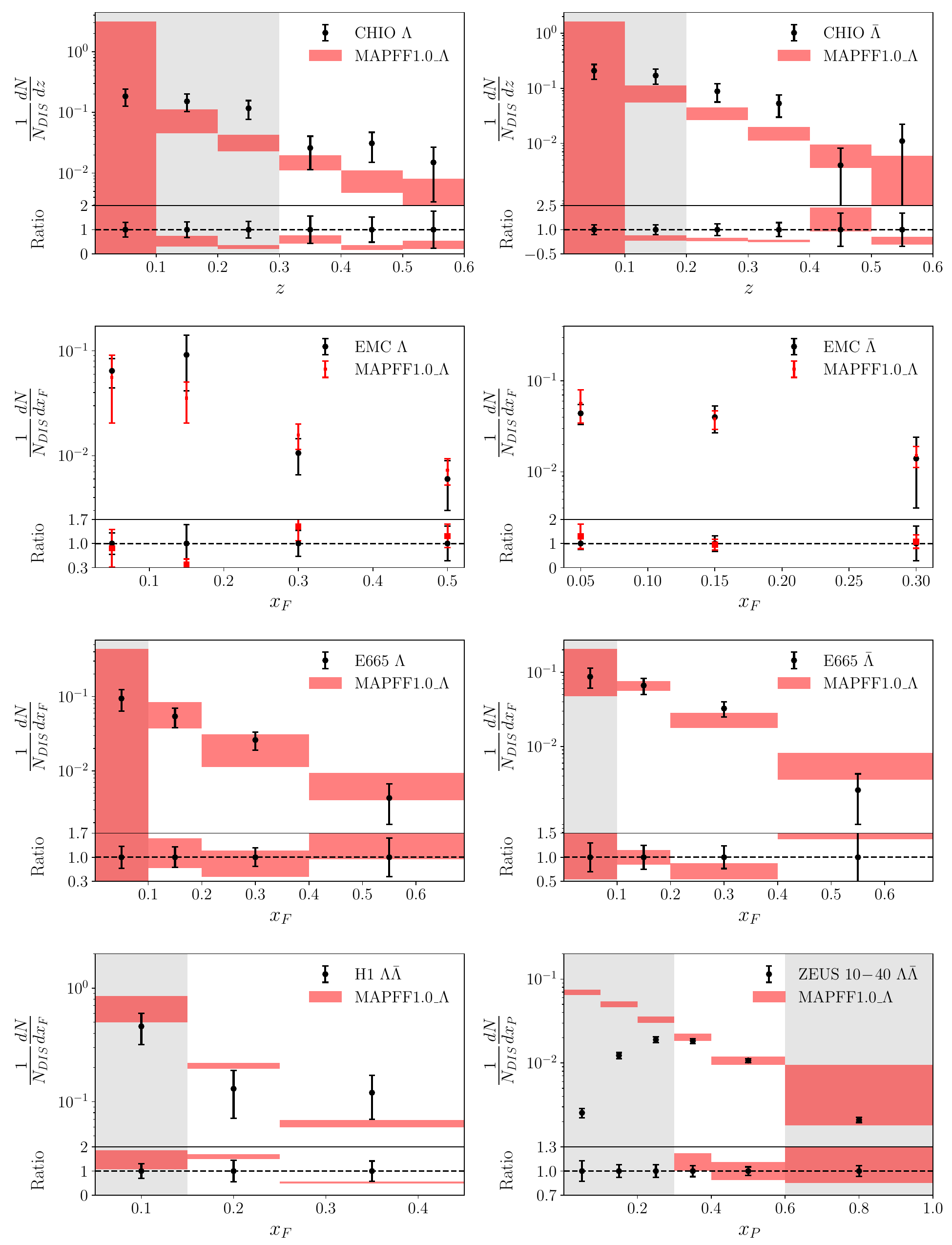}
    \caption{Comparison of selected SIDIS datasets with the \maplambda\ calculation. In the upper panels, the actual experiment-specific observable is plotted, while the lower panels display the corresponding ratio of the \maplambda\  values to the measured ones. Shaded areas in grey at low and/or high values of $z$ represent the regions excluded from the fit due to the imposed kinematic constraints. When the observable of the relative dataset is given for a bin in $z$, the prediction is represented by a band, otherwise it is computed at the average $z$ value given by the experiment. }
    \label{image:sidis_a}
\end{figure}

\begin{figure}
\centering
    \includegraphics[width=0.93\linewidth]{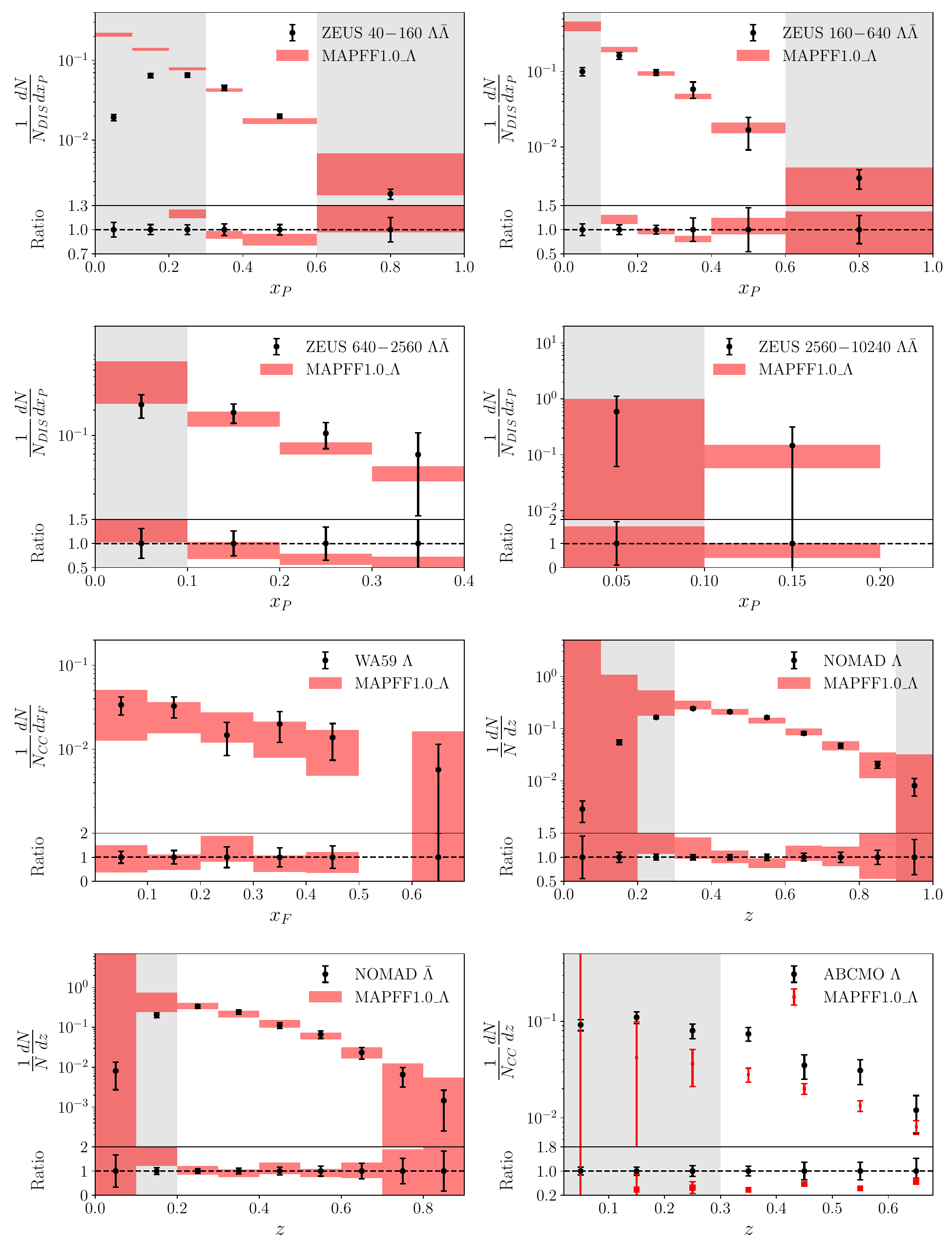}
    \caption{Comparison of selected SIDIS datasets with the \maplambda\ calculation. In the upper panels, the actual experiment-specific observable is plotted, while the lower panels display the corresponding ratio of the \maplambda\  values to the measured ones. Shaded areas in grey at low and/or high values of $z$ represent the regions excluded from the fit due to the imposed kinematic constraints. When the observable of the relative dataset is given for a bin in $z$, the prediction is represented by a band, otherwise it is computed at the average $z$ value given by the experiment.}
    \label{image:sidis_b}
\end{figure}

In this section, we report the comparison at NNLO between data and \maplambda\ calculations for all the datasets included in the analysis, listed in table~\ref{table:sia} for SIA and in table~\ref{table:sidis} for SIDIS, together with the corresponding ratio of the \maplambda\  values to the measured ones. In figures~\ref{image:sia_a},~\ref{image:sia_b}, and~\ref{image:sia_c}, the comparisons for SIA datasets are outlined, whereas SIDIS datasets are collected in figures~\ref{image:sidis_a} and~\ref{image:sidis_b}. Altogether, they constitute the dataset of the {\maplambda} fit.

\bibliographystyle{utphys}
\bibliography{references}
\end{document}